\documentclass[twocolumn]{aastex62}

\usepackage{natbib}
\usepackage{amsmath}	% Advanced maths commands
\usepackage{amssymb}	% Extra maths symbols

\received{}
\revised{}
\accepted{}
\submitjournal{ApJ}

\shorttitle{cSFGs as Old Starbursts Becoming Quiescent}
\shortauthors{G\'omez-Guijarro et al.}

\begin{document}

\title{Compact Star-Forming Galaxies as Old Starbursts Becoming Quiescent}

\correspondingauthor{C. G\'omez-Guijarro}
\email{carlos.gomez@nbi.ku.dk}

\author{C. G\'omez-Guijarro}
\affil{Cosmic Dawn Center (DAWN), Denmark}
\affil{Niels Bohr Institute, University of Copenhagen, Lyngbyvej 2, DK-2100 Copenhagen, Denmark}

\author{G. E. Magdis}
\affil{Cosmic Dawn Center (DAWN), Denmark}
\affil{Niels Bohr Institute, University of Copenhagen, Lyngbyvej 2, DK-2100 Copenhagen, Denmark}
\affil{DTU Space, National Space Institute, Technical University of Denmark, Elektrovej 327, DK-2800 Kgs. Lyngby, Denmark}
\affil{Institute for Astronomy, Astrophysics, Space Applications and Remote Sensing, National Observatory of Athens, GR-15236 Athens, Greece}

\author{F. Valentino}
\affil{Cosmic Dawn Center (DAWN), Denmark}
\affil{Niels Bohr Institute, University of Copenhagen, Lyngbyvej 2, DK-2100 Copenhagen, Denmark}

\author{S. Toft}
\affil{Cosmic Dawn Center (DAWN), Denmark}
\affil{Niels Bohr Institute, University of Copenhagen, Lyngbyvej 2, DK-2100 Copenhagen, Denmark}

\author{A. W. S. Man}
\affil{Dunlap Institute for Astronomy \& Astrophysics, 50 St. George Street, Toronto, ON M5S 3H4, Canada}

\author{R. J. Ivison}
\affil{European Southern Observatory, Karl-Schwarzschild-Stra{\ss}e 2, D-85748 Garching, Germany}
\affil{Institute for Astronomy, University of Edinburgh, Royal Observatory, Blackford Hill, Edinburgh EH9 3HJ, UK}

\author{K. Tisani\'c}
\affil{Department of Physics, Faculty of Science, University of Zagreb, Bijenicka cesta 32, 10000 Zagreb, Croatia}

\author{D. van der Vlugt}
\affil{Leiden Observatory, Leiden University, P.O. Box 9513, 2300 RA Leiden, the Netherlands}

\author{M. Stockmann}
\affil{Cosmic Dawn Center (DAWN), Denmark}
\affil{Niels Bohr Institute, University of Copenhagen, Lyngbyvej 2, DK-2100 Copenhagen, Denmark}

\author{S. Martin-Alvarez}
\affil{Subdepartment of Astrophysics, University of Oxford, Keble Road, Oxford, OX1 3RH, UK}

\author{G. Brammer}
\affil{Cosmic Dawn Center (DAWN), Denmark}
\affil{Niels Bohr Institute, University of Copenhagen, Lyngbyvej 2, DK-2100 Copenhagen, Denmark}

\begin{abstract}

Optically-compact star-forming galaxies (SFGs) have been proposed as immediate progenitors of quiescent galaxies, although their origin and nature are debated. Were they formed in slow secular processes or in rapid merger-driven starbursts? Addressing this question would provide fundamental insight into how quenching occurs. We explore the location of the general population of galaxies with respect to fundamental star-forming and structural relations, identify compact SFGs based on their stellar core densities, and study three diagnostics of the burstiness of star formation: 1) Star formation efficiency, 2) interstellar medium (ISM), and 3) radio emission. The overall distribution of galaxies in the fundamental relations points towards a smooth transition towards quiescence while galaxies grow their stellar cores, although some galaxies suddenly increase their specific star-formation rate when they become compact. From their star formation efficiencies compact and extended SFGs appear similar. In relation to the ISM diagnostic, by studying the CO excitation, the density of the neutral gas, and the strength of the ultraviolet radiation field, compact SFGs resemble galaxies located in the upper envelope of the SFGs main sequence, although yet based on a small sample size. Regarding the radio emission diagnostic we find that galaxies become increasingly compact as the starburst ages, implying that at least some compact SFGs are old starbursts. We suggest that compact SFGs could be starburts winding down and eventually crossing the main sequence towards quiescence.

\end{abstract}


\section{Introduction} \label{sec:intro}

In the past decade, various studies have revealed a tight correlation between the star-formation rate (SFR) and the stellar mass of star-forming galaxies (SFGs). The so-called main sequence (MS) of star formation \citep[e.g.,][]{2004MNRAS.351.1151B,2007ApJ...670..156D,2007A&A...468...33E,2007ApJ...660L..43N,2012ApJ...754L..29W} exhibits a small scatter observed at least up to $z \sim 4$ \citep[$\sim 0.3$\,dex; e.g.,][]{2007A&A...468...33E,2007ApJ...660L..43N,2012ApJ...754L..29W,2014ApJS..214...15S,2015A&A...575A..74S} implying that secular evolution is the dominant mode of stellar growth where gas inflows, outflows, and consumption through star formation are in equilibrium \citep[e.g.,][]{2010ApJ...713..686D,2010MNRAS.407.2091G,2010Natur.463..781T,2013MNRAS.435..999D,2015MNRAS.446.1939F}. Therefore, SFGs spend most of their time evolving as extended star-forming disks. Conversely, quiescent galaxies (QGs), having low specific star formation rate (sSFR), are located below the MS and are typically more compact than SFGs for a fixed stellar mass and redshift \citep[e.g.,][]{2014ApJ...788...28V}. The quenching of star formation and the departure from the MS must imply the build-up of a central stellar core \citep[e.g.,][]{2003MNRAS.346.1055K,2014ApJ...788...11L,2014ApJ...791...45V,2017ApJ...838...19W,2017ApJ...840...47B}.

A population of galaxies have been proposed to be the missing link between the extended SFGs and the more compact QGs, the so-called compact star-forming galaxies \citep[cSFGs; e.g.,][]{2013ApJ...765..104B,2014ApJ...791...52B,2014Natur.513..394N,2014ApJ...780....1W,2015ApJ...813...23V}. cSFGs are typically located within the scatter of the MS, although their origin and nature are still debated. Given the implications of the small scatter of the MS, several studies advocated that extended SFGs within the MS are capable of building up their stellar cores slowly in their secular evolution \citep[e.g.,][]{2013MNRAS.435..999D,2015MNRAS.450.2327Z,2016MNRAS.457.2790T}. However, starburst galaxies (SBs) dominated by a violent episode of star formation typical of gas-rich mergers that move well above the scatter of the MS are also capable of quickly building up compact stellar cores and have been also proposed as early progenitors of QGs \citep[e.g.,][]{2008A&A...482...21C,2010MNRAS.406..230R,2013Natur.498..338F,2013ApJ...772..137I,2014ApJ...782...68T,2017Natur.546..510T,2018ApJ...856..121G}.

Did the build-up of the stellar core, formation of cSFGs, and subsequent quenching of star formation happen as the product of the slow secular evolution or rapidly? Or in other words, is it the natural endpoint of secular galaxy evolution when a sufficiently large bulge has build up or does it require an external event like a merger-induced starburst to compress the gas at the center of the collision and quickly convert it into stars?

Some works have recently pointed towards the starburst nature of cSFGs based on their interstellar medium (ISM) properties \citep[e.g.,][]{2016ApJ...832...19S,2017ApJ...851L..40B,2017A&A...602A..11P,2017ApJ...841L..25T,2018MNRAS.476.3956T}. However, these results are still based on a handful of cSFGs. Other recent works have also indicated the existence of a population of SBs within the scatter of the MS based on their high SFR surface densities based on far-infrared (FIR) and radio observations \citep{2018A&A...616A.110E,2019A&A...625A.114J}, and also, a population of galaxies within the scatter of the MS undergoing compact star formation based on CO lines observations \citep{2019arXiv190502958P}.

In this work we explore the location of extended and compact SFGs and QGs with respect to the MS and structural relations and investigate three diagnostics of the burstiness of star formation: 1) Star formation efficiency (SFE), 2) ISM (CO excitation, density of the neutral gas, and strength of the ultraviolet field), and 3) radio emission (FIR/radio ratio and radio spectral slope). We aim at shedding some light on how rapidly the build-up of compact stellar cores and subsequent quenching of star formation takes place.

The layout of the paper is as follows. We describe the sample selection and identification of extended, compact SFGs, and QGs in Section~\ref{sec:sample}. In Section~\ref{sec:c_sf} we explore the distribution of SFGs and QGs with respect to fundamental star-forming and structural relations, followed by a discussion in Section~\ref{sec:discussion}. We investigate SFE, ISM, and radio emission diagnostics of the burstiness of star formation in Section~\ref{sec:csfgs_mssb}. We summarize the main findings and conclusions in Section~\ref{sec:summary}.

Throughout this work we adopted a concordance cosmology $[\Omega_\Lambda,\Omega_M,h]=[0.7,0.3,0.7]$ and Chabrier initial mass function (IMF) \citep{2003PASP..115..763C}.

\section{Selection of Compact Star-Forming Galaxies} \label{sec:sample}

\subsection{Optical Sample} \label{subsec:opt_sample}

There are several cSFGs selection criteria in the literature. We followed the \citet{2017ApJ...840...47B} identification criteria based on structural and star-forming relations. \citet[][see also references therein]{2017ApJ...840...47B} showed that SFGs and QGs follow distinct trends in the stellar density versus stellar mass plane, with QGs being offset to higher densities at fixed stellar mass and redshift. cSFGs are galaxies that follow the structural relation of QGs, while being star-forming. Therefore, cSFGs are more compact than regular SFG at fixed stellar mass and redshift. \citet{2017ApJ...840...47B} proposed a compactness selection threshold in the core density ($\Sigma_{\rm{1}}, r < 1$\,kpc) as the most efficient way of selecting cSFGs, given the small scatter of the $\Sigma_{\rm{1}}-M_{\rm{*}}$ QGs structural relation and mild normalization decline with redshift. This threshold is: 

\begin{equation}
\Sigma_{\rm{1}} - 0.65(\log M_{\rm{*}} - 10.5) > \log B(z) - 0.2, 
\end{equation}

where $\log B(z)$ have a small redshift dependence ranging between 9.5--9.8\,$M_{\rm{*}}$ kpc$^{-2}$ \citep[see][for details on its definition]{2017ApJ...840...47B}. For simplicity, we will refer to this threshold as $\Sigma_{\rm{1QGs}}$ hereafter. In contrast, the QGs structural relation based on the effective density ($\Sigma_{\rm{e}}, r < r_{\rm{e}}$, where $r_{\rm{e}}$ is the effective radius) would be less efficient identifying cSFGs as it shows larger scatter and variation of the normalization with redshift. By extension, other selection criteria based on stellar mass and effective radius would be also less efficient. Since by construction both cSFGs and QGs follow the same structural relation, the distinction between cSFGs and QGs is based on the distance to the main sequence of star formation ($\Delta \rm{MS}$), defined as the ratio of the sSFR to the sSFR of the MS at the same stellar mass and redshift ($\Delta \rm{MS} = \rm{sSFR/sSFR_{MS}}$). The threshold in \citet{2017ApJ...840...47B} is $\Delta \rm{MS} = -0.7$\,dex, which corresponds to $\sim 2.5\sigma$ below the MS. cSFGs are galaxies above this threshold in star formation.

For our analysis we worked with the cosmological fields COSMOS \citep{2007ApJS..172....1S} and GOODS-North \citep{2003mglh.conf..324D}. As a starting point, we employed the 3D-\textit{HST} survey catalogs \citep{2012ApJS..200...13B,2014ApJS..214...24S,2014ApJ...795..104W,2016ApJS..225...27M} in the CANDELS \citep{2011ApJS..197...35G,2011ApJS..197...36K} portion of COSMOS and GOODS-North, from which we collected stellar masses, SFRs, and redshifts. The structural parameters were gathered from the associated catalogs in \citet{2014ApJ...788...28V}. We trimmed the catalogs following \citet{2017ApJ...840...47B}: 1) $0.5 < z < 3.0$, to guarantee that \citet{2017ApJ...840...47B} structural relations exist; 2) $\log M_{\rm{*}} > 9.0$ for SFGs, $\log M_{\rm{*}} > 10.0$ for QGs, and $H_{\rm{F160W}} < 25.5$, to guarantee that the minimum requirements in the validity of the structural parameters are fulfilled \citep{2012ApJS..203...24V,2014ApJ...788...28V}, where sources flagged as catastrophic failures in the surface brightness profile fits were excluded. \citet{2012ApJS..203...24V,2014ApJ...788...28V} showed that the effective radius ($r_{\rm{e}}$) and S\'ersic index ($n$) have uncertainties $< 10\%$ for galaxies $H_{\rm{F160W}} < 24.5$, and discussed that a redshift-dependent mass threshold of $\log M_{\rm{*}} > 8.5$--9.75 for SFGs and $\log M_{\rm{*}} > 9.0$--10.3 for QGs at $0.5 < z < 3.0$ guarantees that the galaxies are $H_{\rm{F160W}} < 24.5$. \citet{2017ApJ...840...47B} chose $\log M_{\rm{*}} > 9.0$ for SFGs and $\log M_{\rm{*}} > 10.0$ for QGs as a good compromise between dynamical range in stellar mass and accuracy in the structural parameters. To trace approximately the same rest-frame wavelength as a function of redshift we used the structural parameters derived in the $J_{\rm{F125W}}$-band at $z < 1.5$ and $H_{\rm{F160W}}$-band at $z \geq 1.5$. We refer to the sample resulting from this selection as our parent optical sample, which is composed of 13703 galaxies (7222 in COSMOS and 6481 in GOODS-North) with 416 cSFGs (227 in COSMOS and 189 in GOODS-North).

\subsection{Far-Infrared Sample} \label{subsec:fir_sample}

The "super-deblended" FIR to submillimeter photometric catalogs in COSMOS \citep{2018ApJ...864...56J} and GOODS-North \citep{2018ApJ...853..172L} provided fluxes from highly-confused low-resolution data to optical counterparts by using a mix of priors based on high-spatial resolution bands (\textit{Spitzer}/MIPS 24\,$\mu$m, VLA 1.4, and 3\,GHz for COSMOS; \textit{Spitzer}/MIPS 24\,$\mu$m, VLA 1.4\,GHz for GOODS-North). We trimmed these catalogs to sources with a combined signal-to-noise $\rm{S/N}_{\rm{FIR+mm}} \geq 5$ (where $\rm{S/N}_{\rm{FIR+mm}}$ is the quadrature-sum of the S/N in all $\lambda \geq 100$\,$\mu$m bands in the catalogs \citep{2018ApJ...853..172L,2018ApJ...864...56J}). We refer to the sample resulting from this selection as our FIR sample, which is composed of 968 galaxies (357 in COSMOS and 611 in GOODS-North) with 73 cSFGs (26 in COSMOS and 47 in GOODS-North).

\subsubsection{Rayleigh-Jeans and Radio Subsets} \label{subsubsec:rjradio_subset}

We separated a subset of galaxies in the FIR sample that have at least one detection at $\rm{S/N} \geq 3$ above a rest-frame wavelength of 250\,$\mu$m, so-called Rayleigh-Jeans (R-J) side of the FIR spectral energy distribution (SED), required to obtain gas mass estimates (Section~\ref{subsubsec:fir_prop}). This comprises our R-J subset of the FIR sample, composed of 59 galaxies (24 in COSMOS and 35 in GOODS-North) with 5 cSFGs (4 in COSMOS and 1 in GOODS-North).

Additionally, we cross-matched our FIR catalog with radio catalogs from the Giant Metrewave Radio Telescope (GMRT) at 325\,MHz and 610\,MHz in COSMOS \citep{2019A&A...621A.139T} and at 610\,MHz in GOODS-North \citep{2015A&A...573A..45M}. Besides, we substitued the COSMOS "super-deblended" FIR catalog 3\,GHz measurements for those in the COSMOS-XS survey (D. van der Vlugt et al. 2019, in preparation; H. Algera et al. 2019, in preparation) for overlapping sources in both catalogs, given the increased depth of the latter survey. We looked for radio counterparts within the half power beam width (HPBW) at each frequency. We separated a subset of galaxies that have at least two $\rm{S/N} \geq 5$ detections at any available radio frequency (325\,MHz, 610\,MHz, 1.4\,GHz, and 3\,GHz in COSMOS; 610\,MHz, 1.4\,GHz, and 3\,GHz in GOODS-North), required for our radio diagnostic analysis (Section~\ref{subsec:radio}). This comprises our radio subset of the FIR sample, composed of 60 galaxies (23 in COSMOS and 37 in GOODS-North) with 7 cSFGs (2 in COSMOS and 5 in GOODS-North).

\subsubsection{Far-infrared Properties} \label{subsubsec:fir_prop}

We derived infrared luminosities ($L_{\rm{IR}}$) and infrared-based star formation rates ($\rm{SFR_{IR}}$) for our FIR sample. In order to derive these quantities, we first fitted the mid-IR-to-millimeter SED using the \citet{2007ApJ...657..810D} models. These models linearly combine two dust components, one coming from the diffuse ISM and one heated by a power-law distribution of starlight associated with photodissociation regions (PDRs). The methodology was presented in detail in previous studies \citep[e.g.,][]{2012ApJ...760....6M,2017A&A...603A..93M,2016A&A...587A..73B}. We also included an active galactic nuclei component (AGN) to ensure that the derived FIR properties account for star formation only. The best fit to the models were derived through $\chi^{2}$ minimization and the uncertainties were calculated over 1000 realizations of the observed SED perturbing the photometry within the errors. $L_{\rm{IR}}$ was calculated by integrating the best fit to the SED in the range 8--1000\,$\mu$m and $\rm{SFR_{IR}}$ from the $L_{\rm{IR}}$ to $\rm{SFR_{IR}}$ conversion in \citet{1998ARA&A..36..189K} for a Chabrier IMF. One of the parameters derived from the fit is the dust mass ($M_{\rm{dust}}$), which can be used to derive gas masses ($M_{\rm{gas}}$). In order for the $M_{\rm{gas}}$ estimates to be reliable, it is required at least one detection in the R-J side of the SED. Therefore, we derive $M_{\rm{gas}}$ only for our R-J subset of the FIR sample. We used the metallicity-dependent gas-to-dust mass ratio technique ($\delta_{\rm{GD}}$), adopting the $M_{\rm{gas}}/M_{\rm{dust}}$--$Z$ relation of \citet{2012ApJ...760....6M} ($\log (M_{\rm{dust}}/M_{\rm{gas}}) = (10.54 \pm 1.0) - (0.99 \pm 0.12) \times (12 + \log(O/H))$), where the gas-phase metallicity is calibrated using the \citet{2004MNRAS.348L..59P} scale. We assumed a solar metallicity for all galaxies that corresponds to a $M_{\rm{gas}}/M_{\rm{dust}} \sim 90$. Another method to derive $M_{\rm{gas}}$ is the single band measurement of the dust continuum emission flux on the R-J side of the SED \citep[e.g.,][]{2014ApJ...783...84S,2015ApJ...799...96G,2016ApJ...820...83S,2016ApJ...833..112S}. Both the $\delta_{\rm{GD}}$ method and the single-band measurement of the dust emission method from \citet{2016ApJ...820...83S} yielded consistent results on average, with a median and median absolute deviation ratio of $M_{\rm{gas}}^{\rm{GD}}/M_{\rm{gas}}^{\rm{R-J}} = 0.88 \pm 0.41$. In the following we adopt $M_{\rm{gas}}$ estimates from the $\delta_{\rm{GD}}$ method since it employs all datapoints in the SED and, particularly, when there are several detections in the R-J side. Note that both methods account for the total gas budget of the galaxies, including the molecular ($M_{\rm{H_2}}$) and the atomic phases ($M_{\rm{HI}}$).

\subsection{Active Galactic Nuclei Flagging} \label{subsec:agn}

AGN activity is known to correlate with star formation \citep[e.g.,][]{2003MNRAS.346.1055K,2010ApJ...711..284S,2015ApJ...800L..10R} and to be present in a large fraction of cSFGs at $2 < z < 3$ \citep{2014ApJ...791...52B}. We kept track of the galaxies with evidence of AGN activity from several indicators for potential systematics in the AGN population respect to the general population. We flagged all the galaxies for which the AGN fraction from our FIR SED modeling is $\geq 20\%$. In addition, we checked for X-ray bright AGN ($\log L_{\rm{X}} > 42.5$, absorption-corrected soft and hard X-ray luminosity) in the COSMOS \citep[\textit{Chandra} COSMOS Legacy Survey;][]{2016ApJ...819...62C,2016ApJ...817...34M} and GOODS-North \citep{2016ApJS..224...15X} X-ray catalogs. Finally, we identified radio-excess AGN as those having a significantly low FIR/radio ratio ($q < 1.68$) following \citet{2013A&A...549A..59D}. These AGN flagging accounts for unobscured to relatively obscured bright AGN and radio loud AGN, particularly for the FIR sample for which all AGN indicators are available. We found that 22\% of the massive ($\log M_{\rm{*}} \geq 10.3$) cSFGs have an AGN.

\section{Compactness and Star Formation} \label{sec:c_sf}

In this section we explore the location of SFGs and QGs with respect to the MS of star formation and the structural relation of QGs.

For each galaxy in the parent optical sample we calculated $\Delta \rm{MS}$, adopting the MS definition of \citet{2014ApJ...795..104W}, and the distance to the compactness selection threshold in the core density based on the QGs structural relation ($\Delta \Sigma_{\rm{QGs}} = \Sigma_{\rm{1}}/\Sigma_{\rm{1QGs}}$), adopting the relation definition in \citet{2017ApJ...840...47B}, at its stellar mass and redshift. We will refer to extended SFGs as SFGs located at $\Delta \Sigma_{\rm{QGs}} < 1.0$, as opposed to cSFGs located at $\Delta \Sigma_{\rm{QGs}} > 1.0$. Similarly, extended QGs are QGs at $\Delta \Sigma_{\rm{QGs}} < 1.0$ and compact QGs are QGs at $\Delta \Sigma_{\rm{QGs}} > 1.0$.

Note that the SFRs in the 3D-\textit{HST} catalogs are defined as $\rm{SFR_{IR+UV}} = 1.09 \times 10^{-10} (L_{\rm{IR}} + 2.2L_{\rm{UV}})$, where $L_{\rm{IR}}$ is obtained through a conversion from the observed \textit{Spitzer}/MIPS 24\,$\mu$m flux density to $L_{\rm{IR}}$ (8--1000\,$\mu$m) based on a single template. $L_{\rm{UV}}$ is the total integrated rest-frame luminosity in the range 1216--3000\,\AA. For the FIR sample ($\sim 7\%$ of the parent optical sample) we substituted the $\rm{SFR_{IR}}$ contribution for the one we obtained in Section~\ref{subsubsec:fir_prop}, since it uses all the information available in the FIR SED, as opposed to a single template which could dilute galaxies that intrinsically deviate from the template. We checked that making this $\rm{SFR_{IR}}$ substitution does not introduce a systematic bias respect to \citet{2014ApJ...795..104W} MS definition, which could alter our $\Delta \rm{MS}$ values.

\subsection{General Trends} \label{subsec:c_sf_gen}

\begin{figure*}
\begin{center}
\includegraphics[width=0.90\textwidth]{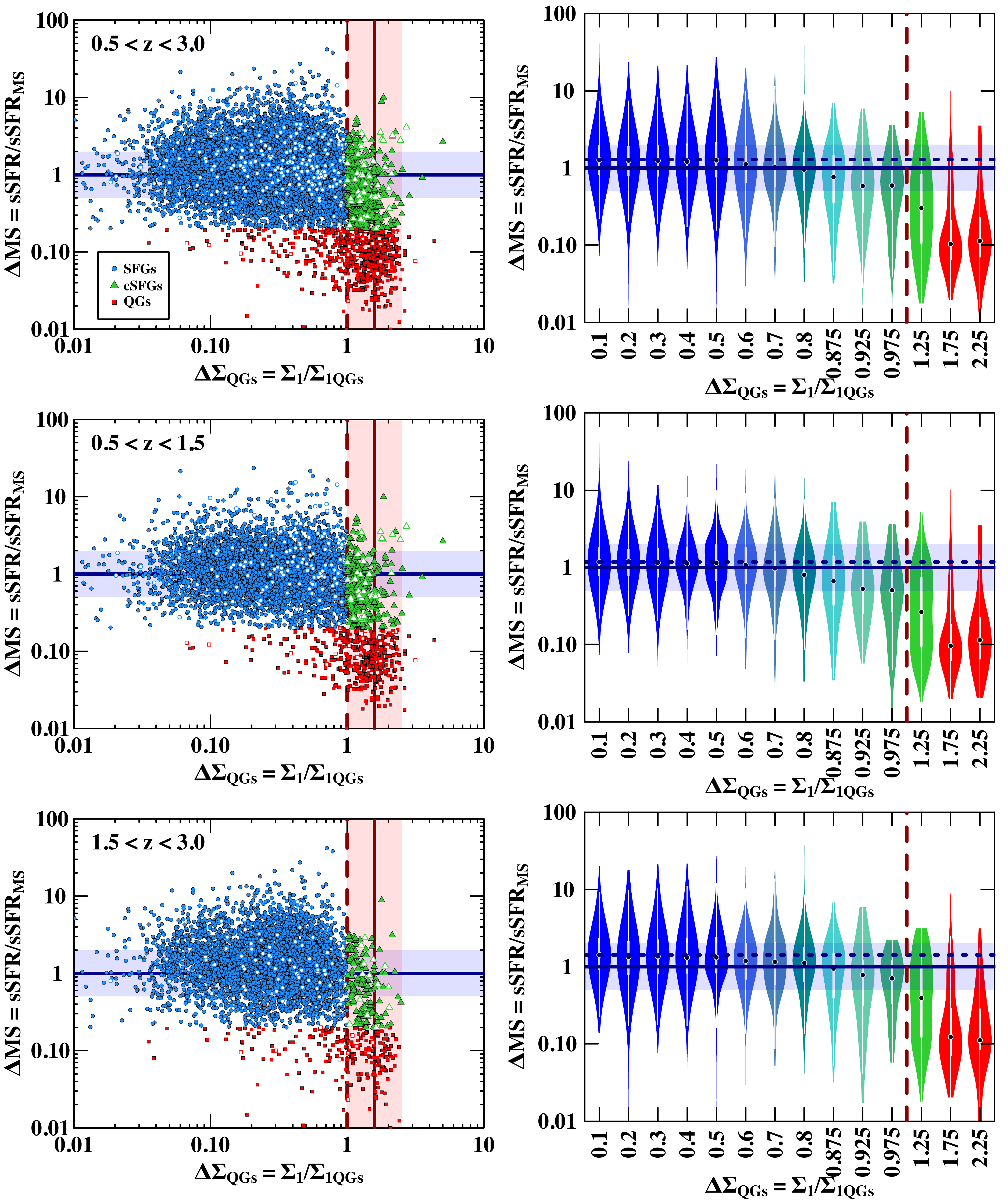}
\caption{$\Delta \rm{MS}$-$\Delta \Sigma_{\rm{QGs}}$ plane. First row left panel: Parent optical sample at all redshifts ($0.5 < z < 3.0$) composed of extended SFGs (blue), cSFGs (green), and QGs (red). AGN-flagged galaxies are represented with empty symbols. Right panel: Violin plot, a combination of a box and whiskers and a density plot. The black dot in the middle is the median value and the thick white bar in the centre represents the interquartile range ($\rm{IQR = Q3 - Q1}$). The thin white line extended from it indicates the upper and lower adjacent values, defined as $\rm{LAV = Q1 - 1.5IQR}$ and $\rm{UAV = Q3 + 1.5IQR}$, respectively. The width of the colored area represents the probability density of a given value in the Y axis. Note that the colors of the violin plot are not directly linked to extended, cSFGs, or QGs, but were chosen to be a representative color of their abundance in each bin. The MS as defined in \citet{2014ApJ...795..104W} is represented with a blue solid line. The 1$\sigma$ scatter of the MS \citep[$\sim 0.3$\,dex; e.g.,][]{2012ApJ...754L..29W,2015A&A...575A..74S} is indicated with a blue shaded region. The median $\Delta \rm{MS} = 1.25$ in the phase up to $\Delta \Sigma_{\rm{QGs}} = 0.5$ is represented as a dashed blue line. The QGs structural relation as defined in \citet{2017ApJ...840...47B} is represented with a red solid line. The compactness selection threshold is represented as a red dashed line. Second row: Similar to the first row for galaxies with $0.5 < z < 1.5$. Third row: Similar to the first row for galaxies at $1.5 < z < 3.0$. The typical uncertainties are 0.15\,dex in the X axis and 0.10\,dex in the Y axis.}
\label{fig:dms_dst}
\end{center}
\end{figure*}

In Figure~\ref{fig:dms_dst} we present the $\Delta \rm{MS}$-$\Delta \Sigma_{\rm{QGs}}$ plane for the parent optical sample. The overall distribution reproduces the L-shape reported in \citet{2017ApJ...840...47B}, with the population of cSFGs forming the knee between extended SFGs and cQGs. This was used as an argument in favour of cSFGs as progenitors of QGs at later times, implying that SFGs become compact before they quench. Note that the majority of QGs are compact QGs.

We also explored the behaviour of the $\Delta \rm{MS}$ (Y axis) per bins of $\Delta \Sigma_{\rm{QGs}}$ (X axis) in Figure~\ref{fig:dms_dst}. In order to do so, we draw a violin plot, a combination of a box and whiskers plot and a density plot to visualize the distribution of the data and its probability density. The violin plot has the advantage of showing not only a discrete median value per bin in the X axis, but also different measurements of the scatter and the actual shape of the distribution of the data in the Y axis.

Overall, galaxies start to transition smoothly towards quiescence, since $\Delta \rm{MS}$ decreases continuously for increasing $\Delta \Sigma_{\rm{QGs}}$. Some extended SFGs quench forming extended QGs as they build up their stellar cores. On the other hand, the sharp transition region at $\Delta \Sigma_{\rm{QGs}} \sim 1.0$ indicates that other galaxies become compact before they quench as reported in \citet{2017ApJ...840...47B}. Some extended SFGs become cSFGs and then compact QGs as they build up their stellar cores. The latter would be a more common track since the majority of QGs are compact QGs. Note that it has to be consider that SFGs do not evolve into QGs at the same epoch (i.e., redshift), but into QGs at later times.

The behavior around the $\Delta \Sigma_{\rm{QGs}} \sim 1.0$ transition threshold presents some remarkable features. In the bin centered at $\Delta \Sigma_{\rm{QGs}} = 1.25$, while the median $\Delta \rm{MS}$ decreases abruptly reflecting the sharp transition region mentioned above, the scatter in $\Delta \rm{MS}$ increases in both directions of the Y axis. Particularly, it is interesting the fact that the upper extreme values increase respect to the previous bin centered at $\Delta \Sigma_{\rm{QGs}} = 0.975$. Even in the bin centered at $\Delta \Sigma_{\rm{QGs}} = 1.75$ the upper extreme values still increase, although they are less frequent than in the previous bin. This indicates that, at least some of the galaxies make the transition by increasing their sSFR and going above the scatter of the MS.

Furthermore, it is also interesting that the median $\Delta \rm{MS}$ stays approximately constant up to $\Delta \Sigma_{\rm{QGs}} = 0.5$ and systematically above the MS ($\Delta \rm{MS} = 1.25$; $\sim 0.1$\,dex). This indicates that, while the MS is dominated by galaxies in the extended phase, there is a contribution from more compact galaxies in transition towards quiescence that affects the overall trend that defines the MS lowering its normalization. Another interesting fact is that extended SFGs above the scatter of the MS are far more numerous than cSFGs above the scatter of the MS.

\subsection{Redshift Dependence} \label{subsec:c_sf_z}

In addition to the $\Delta \rm{MS}$-$\Delta \Sigma_{\rm{QGs}}$ plane for the whole redshift range studied, we also present the results in the redshift bins $0.5 < z < 1.5$ and $1.5 < z < 3.0$ in Figure~\ref{fig:dms_dst}. The general trends are similar at low and high redshift, although there are some important differences. At $1.5 < z < 3.0$ the median $\Delta \rm{MS}$ up to $\Delta \Sigma_{\rm{QGs}} = 0.5$ is higher ($\Delta \rm{MS} = 1.35$) than at $0.5 < z < 1.5$ ($\Delta \rm{MS} = 1.16$). This indicates that the MS is more affected at the high-redshift bin than at the low-redshift bin by galaxies that are already in transition towards quiescence and that lower its normalization.

\subsection{Trends for Massive Galaxies} \label{subsec:c_sf_mstar}

\begin{figure*}
\begin{center}
\includegraphics[width=0.90\textwidth]{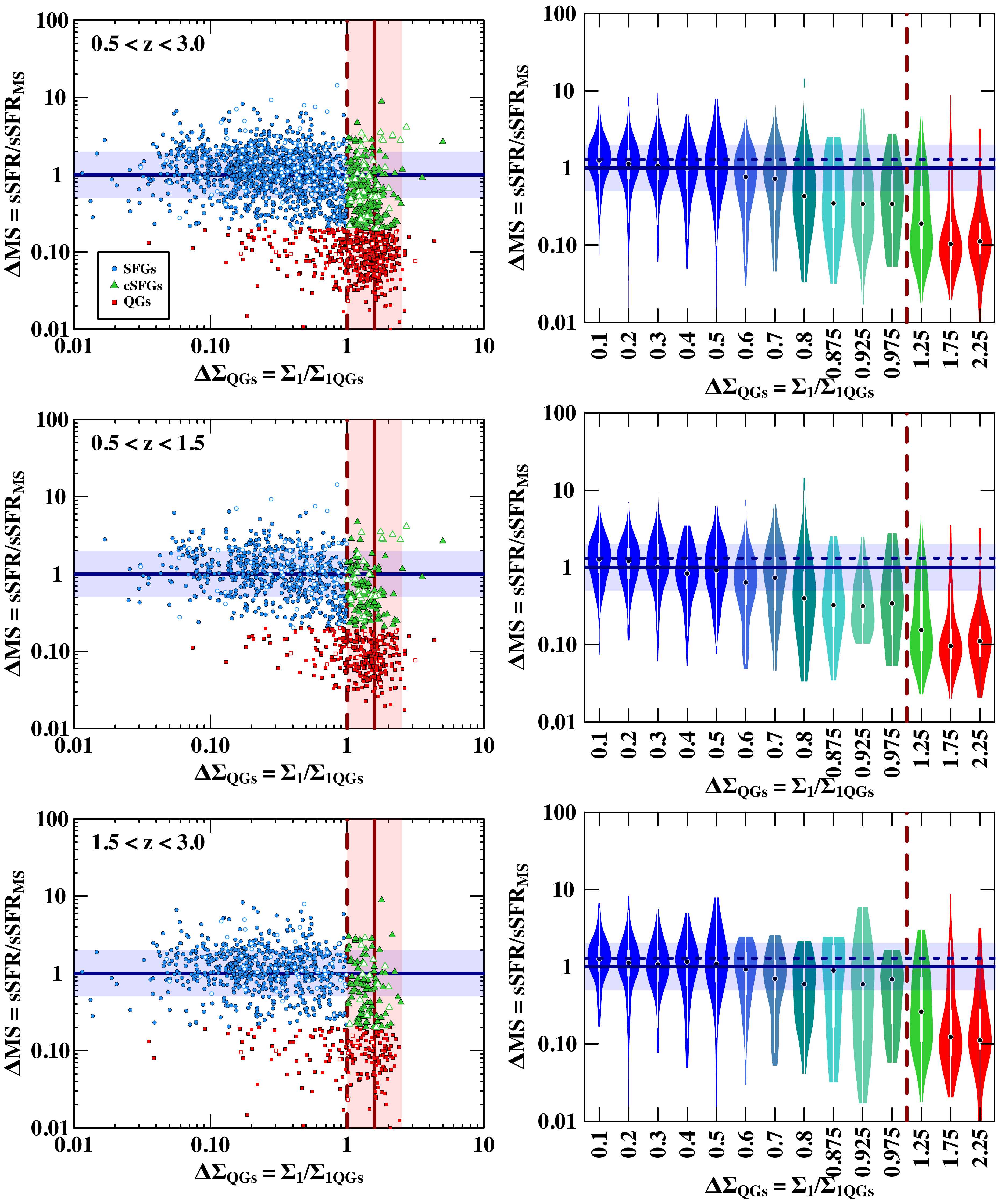}
\caption{$\Delta \rm{MS}$-$\Delta \Sigma_{\rm{QGs}}$ plane as presented in Figure~\ref{fig:dms_dst} for galaxies at $\log M_{\rm{*}} \geq 10.3$.}
\label{fig:dms_dst_mstar}
\end{center}
\end{figure*}

The $\Delta \rm{MS}$-$\Delta \Sigma_{\rm{QGs}}$ planes discussed above follow the selection criteria explained in Section~\ref{subsec:opt_sample}. Particularly, the stellar mass limits are $\log M_{\rm{*}} > 9.0$ for SFGs, $\log M_{\rm{*}} > 10.0$ for QGs. At $\log M_{\rm{*}} \geq 10.3$ the sample is complete for both SFGs and QGs at $ z < 3.0$ \citep{2014ApJ...788...28V} \citep[see also][]{2014ApJS..214...24S,2014ApJ...789..164T,2017ApJ...840...47B}. Therefore, we explored the $\Delta \rm{MS}$-$\Delta \Sigma_{\rm{QGs}}$ plane for the most massive galaxies $\log M_{\rm{*}} \geq 10.3$ in Figure~\ref{fig:dms_dst_mstar}.

Overall, the trends are similar than those discussed in the previous section. Note that the sample statistics are smaller in this case, which has to be taken in consideration when interpreting the plots. One important difference is that the median $\Delta \rm{MS}$ is not approximately constant up to $\Delta \Sigma_{\rm{QGs}} = 0.5$ anymore, but it rather starts to decay since the first bin centered at $\Delta \Sigma_{\rm{QGs}} = 0.1$. This is expected as a consequence of massive galaxies being more dominated by galaxies that are already in transition towards quiescence than low-mass galaxies. In this case we do not appreciate differences in the $\Delta \rm{MS}$ at $\Delta \Sigma_{\rm{QGs}} = 0.1$ at low redshift ($\Delta \rm{MS} = 1.27$) and high redshift ($\Delta \rm{MS} = 1.25$), which indicates that the trend in Section~\ref{sec:c_sf} was dominated by low-mass galaxies. Another important difference is that the number of galaxies above the scatter of the MS respect to those within the scatter of the MS is smaller for massive galaxies. Besides, outliers are less strong (i.e., smaller $\Delta \rm{MS}$), as expected given that for the same increase in SFR the effect in sSFR is smaller as galaxies become more massive.

\section{Are Compact Star-Forming Galaxies Normal Star-Forming Galaxies or Starbursts?} \label{sec:csfgs_mssb}

cSFGs have been proposed as a transition population between being star-forming and quiescence \citep[e.g.,][]{2013ApJ...765..104B,2014ApJ...791...52B,2014Natur.513..394N,2014ApJ...780....1W,2015ApJ...813...23V}. Revealing their nature implies revealing whether the transition to quiescence occurred secularly or rapidly. Phases of abrupt changes in increasing sSFR are typical of SBs. The time a galaxy is detectable in such phase is short, since these are short-lived \citep[e.g.,][]{2008ApJ...680..246T,2008A&A...492...31D}. This means that the number of detectable SBs is small compared to the general population, but the phase can still be very relevant in terms of stellar mass assembly. We examined three diagnostics of the burstiness of star formation: 1) SFE, 2) ISM, and 3) radio emission. The aim is exploring whether cSFGs can be considered normal SFGs, pointing to a more secular evolution, or SBs, pointing to a more rapid evolution. We refer to normal SFGs as those that obey the general trends of SFGs in each diagnostic. Conversely, SBs are outliers to these general trends in each diagnostic (see diagnostics Sections for a more detailed explanation). These definitions of normal SFGs and SBs are independent of their position with respect to the MS. The latter would be a consequence of the physical mechanisms in place (related to our diagnostics) and the effect of these mechanisms on the integrated properties of the galaxies at the time of observation. Note also that the galaxies selected for each of the three diagnostics are not the same sources as the selections do not overlap.

\subsection{Diagnostic 1: Star Formation Efficiency} \label{subsec:sfe}

The star formation law or Kennicutt-Schmidt relation \citep[KS relation;][]{1959ApJ...129..243S,1998ARA&A..36..189K} relates the gas mass and the SFR of SFGs (originally defined using surface densities). Several studies indicated that normal SFGs and SBs follow different trends. SBs have higher SFR per unit of gas mass and, thus, higher star formation efficiencies ($\rm{SFE} = \rm{SFR} / M_{\rm{gas}}$) than normal SFGs \citep[e.g.,][]{2010ApJ...714L.118D,2010MNRAS.407.2091G}. This distinction in SFE serves to distinguish normal SFGs from SBs, regardless of their location with respect to the MS. In this section we apply this SFE-based definition of normal SFGs and SBs.

\begin{figure*}
\begin{center}
\includegraphics[width=\textwidth]{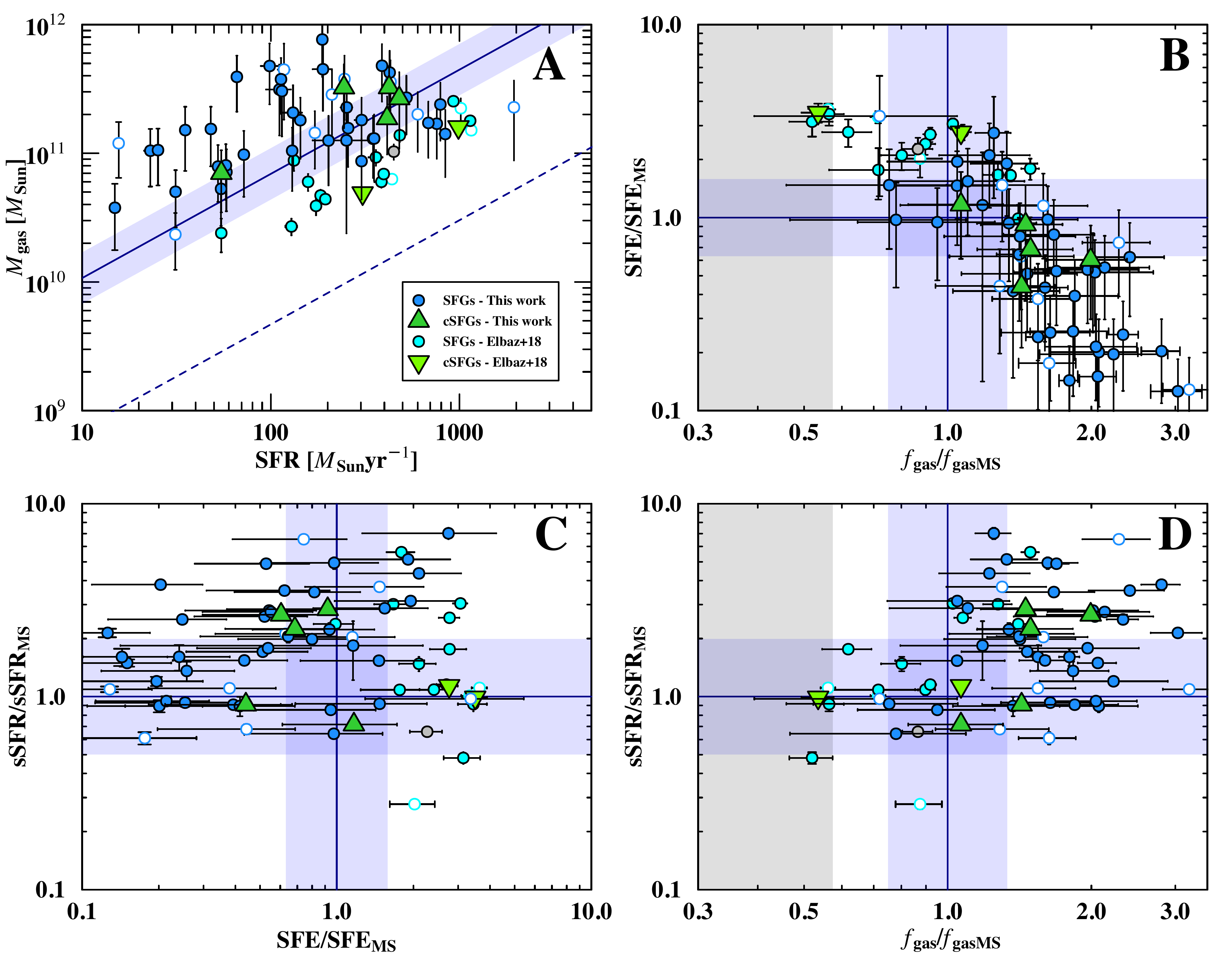}
\caption{Panel A: $M_{\rm{gas}}$-$\rm{SFR}$ plane. Trends for normal SFGs (solid line), with its 0.2\,dex scatter, and SBs (dashed line) from \citet{2014ApJ...793...19S} are shown as reference. Panel B: $\rm{SFE}$-$f_{\rm{gas}}$ plane. Panel C: $\Delta \rm{MS}$-SFE plane. Panel D: $\Delta \rm{MS}$-$f_{\rm{gas}}$ plane. SFE and $f_{\rm{gas}}$ are normalized to the normal SFGs trends in \citet{2014ApJ...793...19S}. The 1$\sigma$ scatter of the MS \citep[$\sim 0.3$\,dex; e.g.,][]{2012ApJ...754L..29W,2015A&A...575A..74S} and the normal SFGs SFE \citep[$\sim 0.2$\,dex;][]{2014ApJ...793...19S} and $f_{\rm{gas}}$ \citep[$\sim 0.125$\,dex;][]{2014ApJ...793...19S} trends are represented as a blue shaded region. Our sample and \citet{2018A&A...616A.110E} sample are shown, classified as extended SFGs (blue circles) or cSFGs (green triangles), except for one of \citet{2018A&A...616A.110E} galaxies (gray circle) unclassifiable due to bad structural parameters. AGN-flagged galaxies are represented with empty symbols. In panels B and D the most favorable $f_{\rm{gas}}$ limit in our selection is shown as a gray shaded region as a reference of the detection threshold (see Section~\ref{sec:discussion}).}
\label{fig:sfe}
\end{center}
\end{figure*}

In Figure~\ref{fig:sfe} we present the locus of our R-J subset of the FIR sample in the $M_{\rm{gas}}$-$\rm{SFR}$ plane in relation with the trends for normal SFGs and SBs in \citet{2014ApJ...793...19S}, where the normal SFGs trend comes from massive MS galaxies in \citet{2014ApJ...793...19S}. Note that we only included the most massive subset of galaxies with $\log M_{\rm{*}} \geq 10.3$, to guarantee that the assumption of solar metallicity to derive $M_{\rm{gas}}$ is valid. We also included \citet{2018A&A...616A.110E} sample, calculating their $\Delta \Sigma_{\rm{QGs}}$ and assessing whether they are extended SFGs or cSFGs, according to \citet{2017ApJ...840...47B} criterion. This can be successfully done for 18/19 galaxies in \citet{2018A&A...616A.110E} as one of the galaxies has bad structural parameters in \citet{2014ApJ...788...28V} catalogs (shown with a gray symbol in Figure~\ref{fig:sfe}). In Figure~\ref{fig:sfe} we also explore the relation between SFE, gas fraction ($f_{\rm{gas}} = M_{\rm{gas}} / (M_{\rm{*}} + M_{\rm{gas}})$), and $\Delta \rm{MS}$, being SFE and $f_{\rm{gas}}$ normalized to the normal SFGs trends as defined by the MS galaxies trends in \citet{2014ApJ...793...19S}.

cSFGs in our sample are consistent with the SFE trend established for normal SFGs. Besides, the extended SFGs in our sample follow the normal SFGs SFE trend as well. Our sample is located within and above the scatter of the MS. The SFE-$f_{\rm{gas}}$ plane exhibits a tendency, being galaxies with lower SFE those with higher $f_{\rm{gas}}$, as expected for galaxies that decrease their SFE as a consequence of increasing their gas content. On the other hand, \citet{2018A&A...616A.110E} sample occupies complementary regions in these diagrams respect to our sample, exhibiting enhanced SFE closer to those of the SBs trend due to low gas fractions. Although beyond the scope of our work, the $\Delta \rm{MS}$-SFE and $\Delta \rm{MS}$-$f_{\rm{gas}}$ planes reflect the general trends of incresing SFE and $f_{\rm{gas}}$ with $\Delta \rm{MS}$ reported in the literature \citep[e.g.,][]{2017ApJ...837..150S,2018ApJ...853..179T}.

These diagrams indicate that our cSFGs are consistent with the trends of normal SFGs with no evidence of SB-like SFE. The combination of our sample with \citet{2018A&A...616A.110E} sample indicates that there is no difference between cSFGs and extended SFGs in terms of their SFE, since both occupy the same regions in the SFE and $f_{\rm{gas}}$ parameter space.

Overall, these results based on the relations among $M_{\rm{gas}}$, stellar mass, and SFR, point towards both secular (lower SFE) and rapid (higher SFE) evolution processes are able to generate cSFGs.

\subsection{Diagnostic 2: Interstellar Medium} \label{subsec:ism}

\subsubsection{CO Excitation} \label{subsubsec:ism_co}

The properties of the ISM are a critical piece of information to study how star formation occurs. The excitation of the CO emission is a good tracer of the ISM properties. It is measured through the line luminosity ratio of CO lines with different rotational number ($J$). The CO\,$(5-4)$/CO\,$(2-1)$ ratio shows the biggest discrepancy between normal SFGs and SBs excitation conditions than any other pair of CO transitions calibrated in the literature \citep[e.g.,][]{2013MNRAS.429.3047B,2015A&A...577A..46D}. \citet{2015A&A...577A..46D} established a benchmark for the excitation conditions of normal SFGs by studying a sample of $BzK$-selected SFGs at $z \sim 1.5$ located within the scatter of the MS. They found that, while less excited than typical SBs such as local ULIRGs or high-redshift SMGs, the average excitation was higher than in the Milky Way. The authors argued that the excitation correlates with the star formation surface density. This, along with the fact that the excitation varied within the sample, motivated us to study whether some of \citet{2015A&A...577A..46D} galaxies are cSFGs. In this section we refer to normal SFGs as those consistent with the CO spectral line energy distribution (SLED) of \citet{2015A&A...577A..46D} $BzK$-selected MS galaxies, while we refer to SBs as those consistent with the median CO SLED of SMGs from \citet{2013MNRAS.429.3047B}.

We cross-matched our parent optical sample with the galaxies in \citet{2015A&A...577A..46D}. We found three of our galaxies, namely GN2359, GN20044, and GN23304, which correspond to $BzK$-4171, $BzK$-16000, and $BzK$-17999 in \citet{2015A&A...577A..46D}, respectively. The missing galaxy $BzK$-21000 in \citet{2015A&A...577A..46D} corresponds to GN38099. Its structural parameters are poorly constrained and, thus, it was excluded from our sample. For the analysis in this section we added it back bearing in mind this caveat. In Figure~\ref{fig:ism} we show the SLEDs in \citet{2015A&A...577A..46D} for these four galaxies. In Table~\ref{tab:ism} we present their $\Delta \rm{MS}$ and $\Delta \Sigma_{\rm{QGs}}$ values.

All four galaxies are extended SFGs and not cSFGs. However, we see that the three galaxies with the highest $\Delta \rm{MS}$ are the ones with the highest CO excitation (GN2359, GN23304, and GN38099), while the other galaxy located right on the MS is the one with the lowest CO excitation (GN20044).

The galaxies with the highest CO excitation are also those with the highest star formation surface density according to \citet{2015A&A...577A..46D}, suggesting that the scatter at higher $\Delta \rm{MS}$ is linked to galaxies progressively forming compact cores.

\subsubsection{Photodissociation Regions} \label{subsubsec:ism_pdr}

Another way of studying the ISM properties is through characterizing the emission from photodissociation regions (PDRs). PDRs are neutral gas regions dominated by far-ultraviolet (FUV) photons. PDR modeling have been used to characterize the strength of the ultraviolet radiation field ($G$) and the density of the neutral gas ($n$) \citep[e.g.,][]{2013MNRAS.435.1493A,2017A&A...602A..11P}. In particular, \citet{2017A&A...602A..11P} employed it to characterize the ISM properties of a cSFGs at $z = 2.225$ located within the scatter of the MS (namely GS30274). They found that the galaxy has SB-like ISM properties, low gas fraction, and high efficiency compared to normal SFGs. The authors interpreted that a previous merger event triggered a central starburst that is quickly building up a dense core of stars responsible for the compact distribution of stellar light. We studied whether galaxies in our sample are similar. In this section we refer to normal SFGs as those consistent with the location of the sample of MS galaxies from \citet{2001ApJ...561..766M} in the $G$-$n$ plane, while we refer to SBs as those consistent with the location of the sample of local ULIRGS from \citet{2003ApJ...597..907D} in the $G$-$n$ plane, identical to the definition used in \citet{2013MNRAS.435.1493A,2017A&A...602A..11P}.

\begin{figure*}
\begin{center}
\includegraphics[width=\textwidth]{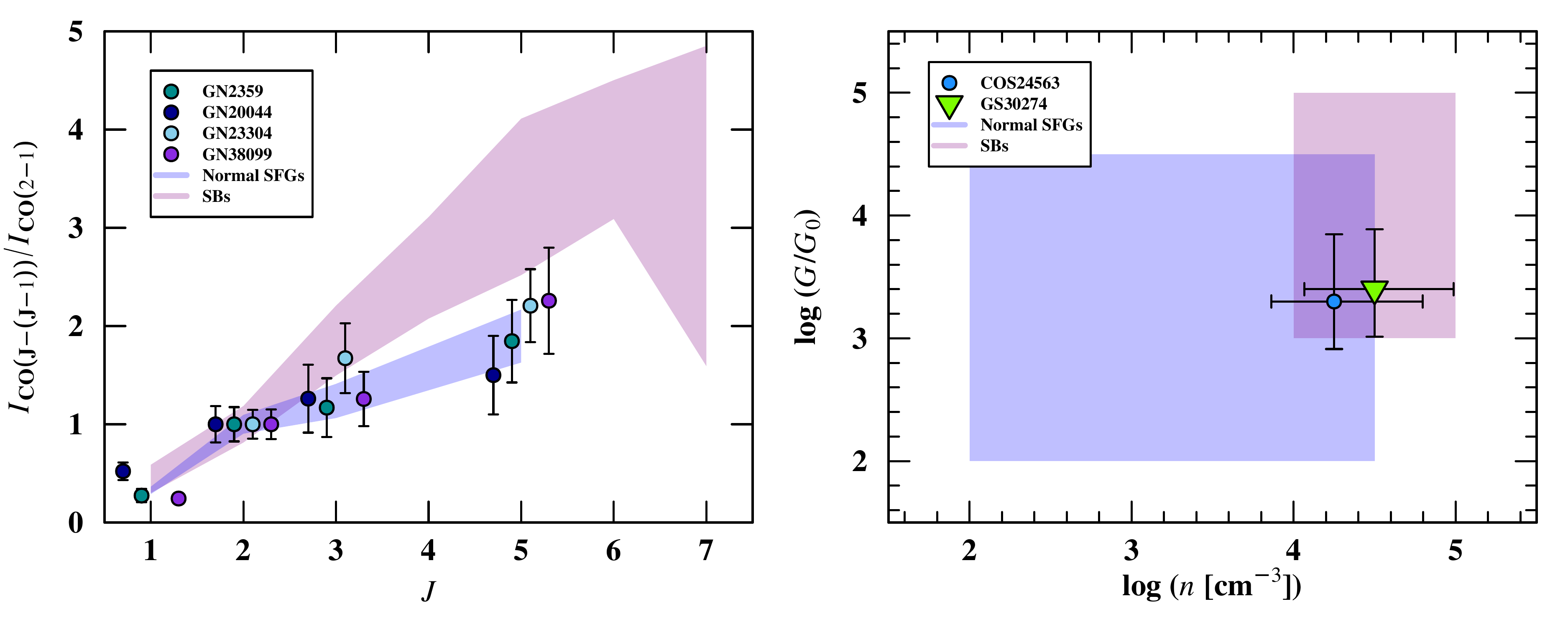}
\caption{Left panel: CO SLEDs from \citet{2015A&A...577A..46D} for the galaxies in our parent optical sample namely GN2359, GN20044, GN23304, and GN38099 which correspond to $BzK$-4171, $BzK$-16000, $BzK$-17999, and $BzK$-21000 in \citet{2015A&A...577A..46D}. The normal SFGs CO SLED is represented as a blue shaded region, corresponding to the average SLED of the sample from \citet{2015A&A...577A..46D}. The SBs CO SLED is represented as a purple shaded region and corresponds to the median SLED of SMGs from \citet{2013MNRAS.429.3047B}. Right panel: $G$-$n$ plane (strength of the ultraviolet radiation field versus density of the neutral gas) with our PDR-modeled COS24563 (18538 in \citet{2018ApJ...869...27V}) and GS30274 (studied in \citet{2017A&A...602A..11P}). The normal SFGs span values within the blue shaded region, which corresponds to the sample of local MS galaxies from \citet{2001ApJ...561..766M}. The SBs span values within the purple shaded regions, which refers to the sample of local ULIRGS from \citet{2003ApJ...597..907D}.}
\label{fig:ism}
\end{center}
\end{figure*}

\citet{2018ApJ...869...27V} presented a survey of atomic carbon [\ion{C}{1}] of FIR-selected galaxies on the MS at $z \sim 1.2$. We cross-matched our parent optical sample with the galaxies in \citet{2018ApJ...869...27V}. We found that one of our extended SFGs was observed in that survey, namely COS24563 (which corresponds to 18538 in \citet{2018ApJ...869...27V}). We performed PDR modeling for this galaxy and also for GS30274 in \citet{2017A&A...602A..11P} for consistency in the methodology and to avoid systematics in the comparison. Besides, we calculated GS30274 $\Delta \Sigma_{\rm{QGs}}$ and checked that it is a cSFGs according to \citet{2017ApJ...840...47B} criterion. Note that the structural parameters of GS30274 are poorly constrained, as also mentioned in \citet{2017A&A...602A..11P}, and do not meet the same quality criteria applied to our sample. In Figure~\ref{fig:ism} we locate the two modeled galaxies in the the $G$-$n$ plane. In Table~\ref{tab:ism} we present the $\Delta \rm{MS}$ and $\Delta \Sigma_{\rm{QGs}}$ values for them.

We estimated the density $n$ (in $\rm{cm^{-3}}$) and the strength of the FUV ($6\,eV < h\nu < 13.6\,eV$) radiation field $G$ (in the Habing field units, $G_0 = 1.6 \times 10^{-3}$\,erg cm$^{-2}$ s$^{-1}$) by comparing the available line luminosities with the 1D modeling of the PDRs by \citet{1999ApJ...527..795K,2006ApJ...644..283K}. This modeling provides a simplified picture of the complex cold ISM phases and their interplay in high-redshift galaxies, but it is enough to capture the average properties of these unresolved systems, without introducing a large number of parameters that cannot be observationally constrained at the current stage. We downloaded the relevant line intensity maps from the online \textsc{PDR Toolbox} \citep{2008ASPC..394..654P}, originally spanning a density interval of $1 < \log n[\rm{cm^{-3}}] < 7$ and FUV intensity range of $-0.5 < \log (G/G_0) < 6.5$ and we resampled them to a 0.05\,dex step grid. We then compared the models and the observations finding the combination of ($n$,$G$) that minimizes the $\chi^2$. We estimated the uncertainties on the best fit ($n$,$G$) both applying the criterion described in \citet{1976ApJ...210..642A} and bootstrapping 1000 times the line luminosities and computing the 68\%, 90\%, and 95\% confidence intervals as inter-percentile ranges. In this work we modeled the neutral atomic carbon $^3P_1\rightarrow^3P_0$ transition ([\ion{C}{1}]$(^3P_1-^3P_0)$, $\nu_{\rm{rest}} = 492.161$\,GHz), a mid-$J$ CO line (CO\,$(4-3)$ or CO\,$(5-4)$ at $\nu_{\rm{rest}} = 461.0408$ and $576.2679$\,GHz, respectively), and the total infrared luminosity ($L_{\rm{IR}}$) removing the possible AGN contribution due to the dusty torus around the central supermassive black hole (see Section~\ref{subsubsec:fir_prop}). The [\ion{C}{1}]$(^3P_1-^3P_0)$/mid-$J$ ratio is primarily sensitive to the density. The use of CO\,$(4-3)$ or CO\,$(5-4)$ as the CO mid-$J$ transition does not affect the results on the density (F. Valentino et al. 2019, in preparation). [\ion{C}{1}]$(^3P_1-^3P_0)$/$L_{\rm{IR}}$ depends on $G$ by construction ($G \propto L_{\rm{IR}}$, \citet{1999ApJ...527..795K}). We thus have roughly perpendicular tracks to determine both ($n$,$G$) parameters \citep[e.g.,][]{2013MNRAS.435.1493A,2017A&A...602A..11P}.

\begin{deluxetable}{lcc}
\tabletypesize{\scriptsize}
\tablecaption{$\Delta \rm{MS}$ and $\Delta \Sigma_{\rm{QGs}}$ for Galaxies in Section~\ref{subsec:ism}} \label{tab:ism}
\tablehead{\colhead{Name} & \colhead{$\Delta \rm{MS}$} & \colhead{$\Delta \Sigma_{\rm{QGs}}$} \\
\colhead{} & \colhead{} & \colhead{}}
\startdata
COS24563 & 3.15 $\pm$ 0.12 & 0.325 $\pm$ 0.011 \\
GN2359   & 2.81 $\pm$ 0.11 & 0.227 $\pm$ 0.012 \\
GN20044  & 1.13 $\pm$ 0.08 & 0.394 $\pm$ 0.007 \\
GN23304  & 2.37 $\pm$ 0.05 & 0.227 $\pm$ 0.012 \\
GN38099  & 5.13 $\pm$ 0.10 & 0.095 $\pm$ 0.003 \\
GS30274  & 2.02 $\pm$ 0.10 & 1.132 $\pm$ 0.011 \\
\enddata
\end{deluxetable}

COS24563 and GS30274 have similar properties as seen in Figure~\ref{fig:ism}. They are both located in the intersection between normal SFGs and SBs ISM properties. COS24563 is placed at a $\Delta \rm{MS}$ slightly above the scatter of the MS, suggesting that the scatter at higher $\Delta \rm{MS}$ is linked to galaxies progressively forming compact cores, similar to the interpretation drawn from the CO excitation.

Overall, the ISM properties from both the CO excitation and PDR modeling suggest that extended SFGs located slightly above the MS (upper-MS galaxies) are capable of hosting an ISM that appears mildly excited and dense similar to the lower envelope of SB-like ISM properties. This suggest that the build-up of a compact core leading to cSFGs could happen secularly, or at least that if coming from the product of a rapid starburst-like event the latent ISM has similar properties to that of upper-MS normal SFGs. However, the small sample sizes and the lack of ISM characterization of cSFGs in this work and over the literature makes it is still difficult to conclude whether the shown ISM properties are the product of a slow secular evolution or the final stages of SBs pointing towards a more rapid evolution.

\subsection{Diagnostic 3: Radio Emission} \label{subsec:radio}

The FIR/radio correlation \citep[FRC; e.g.,][]{1985A&A...147L...6D,1985ApJ...298L...7H,1992ARA&A..30..575C} arises because massive stars ($M_{*} > 8$\,$M_{\odot}$) are responsible for producing ultraviolet photons that are absorbed and re-emitted by dust at FIR wavelengths, and also responsible for accelerating cosmic ray electrons after exploding as supernovae that produces the synchrotron emission at radio wavelengths. \citet{2002A&A...392..377B} modeled the FIR and radio emission of SBs, studying the interplay between the two with the age of the starburst episode and their effect on the FIR/radio ratio ($q \propto L_{\rm{IR}}/L_{\rm{radio}}$) and the slope of the power law radio spectrum ($S \propto \nu^{-\alpha}$), introducing the $q_{\rm{1.4GHz}}$-$\alpha$ diagram as a diagnostic of SBs age evolution. During the very early phase after the star formation ignites, SBs are dominated by FIR emission since only thermal free-free emission from HII regions contributes to the radio emission. At this stage the radio slope is shallow ($\alpha \sim 0.2$) and the radio output is low compared to FIR ($q_{\rm{1.4GHz}} \sim 3$). Then, core-collapse supernovae explosions occur, feeding relativistic electrons to the galactic magnetic fields, and non-thermal synchrotron emission steepens the radio spectrum increasing at the same time the radio output. At this stage the radio slope progressively gets similar to the value typical of normal SFGs ($\alpha \sim 0.8$) and the FIR/radio ratio progressively decreases to a minimum value ($q_{\rm{1.4GHz}} \sim 1.7$). At older ages, the FIR/radio ratio increases again at almost constant radio slope ($\alpha$ and $q_{\rm{1.4GHz}}$ reach asymptotic values). These models were first observationally tested by \citet{2014MNRAS.442..577T} for 870\,$\mu$m-selected SMGs. The authors found that the data populated the predicted region of the parameter space and the stellar masses tend to increase along the SBs evolutionary tracks in the $q_{\rm{1.4GHz}}$-$\alpha$ diagram. We explored the location of our galaxies in this diagram as another diagnostic of their nature. In this section we refer to normal SFGs to the typical SFGs average values of $\alpha = 0.80 \pm 0.25$ \citep[e.g.,][]{1992ARA&A..30..575C,2009MNRAS.397..281I,2010MNRAS.401L..53I} and $q_{\rm{1.4GHz}} = 2.34 \pm 0.26$  in the local universe \citep[e.g.,][]{2001ApJ...554..803Y}.

We calculated $\alpha$ and $q_{\rm{1.4GHz}}$ for the galaxies in the radio subset of the FIR sample. $\alpha$ was obtained through fitting a single power law to the data ($\chi^{2}$ minimization). This corresponds to the slope in the range 325\,MHz--3\,GHz for galaxies in COSMOS and 610\,MHz--1.4\,GHz for galaxies in GOODS-North. The FIR/ratio at 1.4GHz is defined as: 

\begin{equation}
q_{\rm{1.4GHz}} = \log \frac{L_{\rm{IR}}\,\rm{[W]}}{3.75 \times 10^{12}\,\rm{[Hz]}} - \log (L_{\rm{1.4GHz}}\,\rm{[W\,Hz^{-1}]}), 
\end{equation}

\begin{figure*}
\begin{center}
\includegraphics[width=\textwidth]{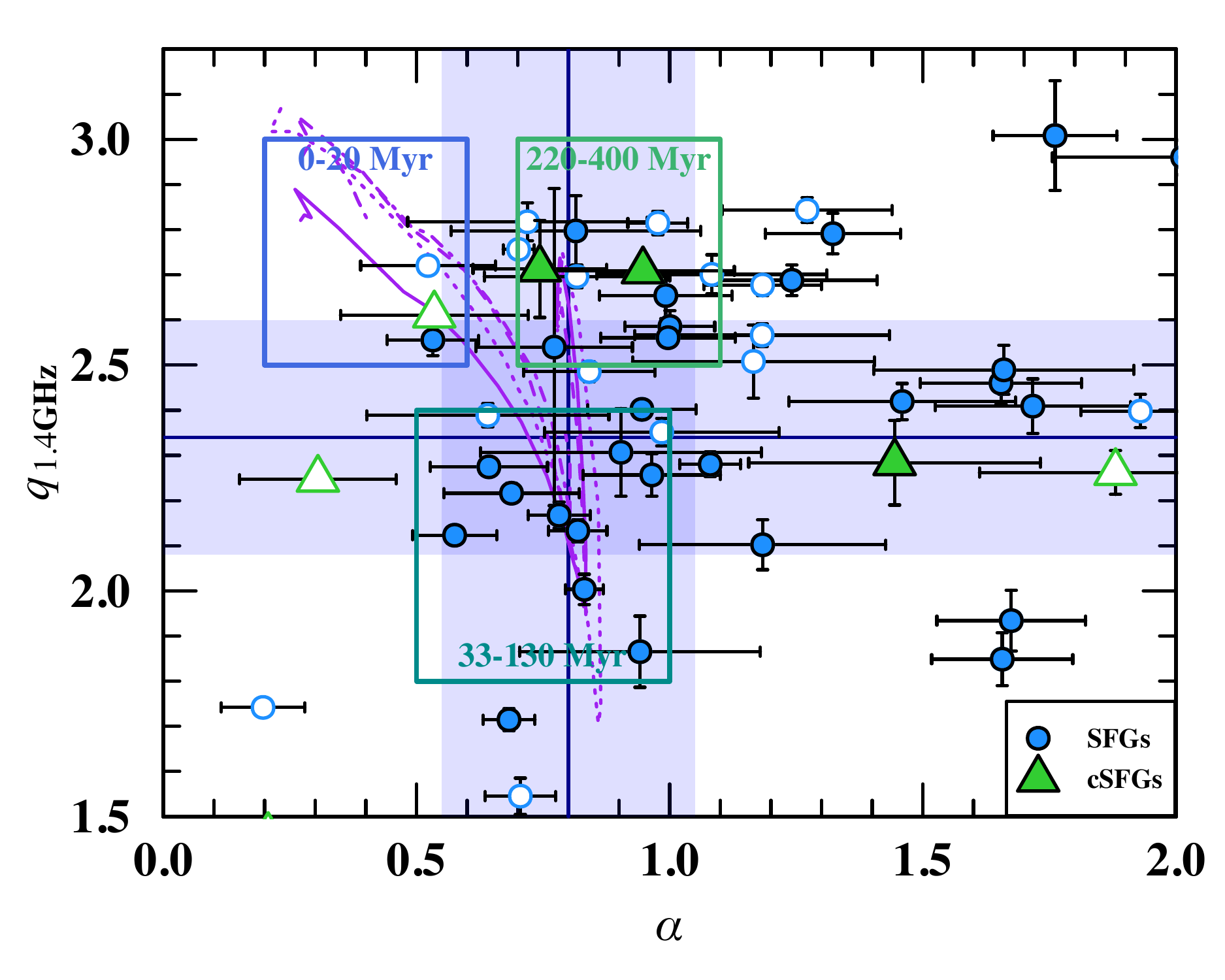}
\caption{$q_{\rm{1.4GHz}}$-$\alpha$ plane. The 1$\sigma$ scatter of the normal SFGs $q_{\rm{1.4GHz}}$ \citep[0.26\,dex;][]{2001ApJ...554..803Y} and $\alpha$ ($\sim$ 0.25\,dex) typical values are represented as a blue shaded region. \citep{2002A&A...392..377B} evolutionary tracks for SBs are plotted as purple lines (different linetypes are examples of different $t_{\rm{burst}}$ in the models). The models run in a logarithmic step ranging $\log t \rm{[yr]} = 6.3$--8.6. Following \citep{2014MNRAS.442..577T} we plot three boxes where young (0--20\,Myr), middle-aged (33--130\,Myr), and old (220--400\,Myr) SBs are expected to be located. AGN-flagged galaxies are represented with empty symbols. Note that some AGN-flagged sources fall outside the plotted region, but we zoom in the area of interest.}
\label{fig:radio}
\end{center}
\end{figure*}

\citep[e.g.,][]{1985ApJ...298L...7H,2001ApJ...554..803Y,2015A&A...573A..45M}, where $L_{\rm{1.4GHz}}$ was calculated via the single power law to the data. The uncertainties were obtained over 10000 realizations of the observed radio SED perturbing the photometry within the errors. In Figure~\ref{fig:radio} we present the locus of our radio subset of the FIR sample in the $q_{\rm{1.4GHz}}$-$\alpha$ plane along with the SBs evolutionary tracks from \citet{2002A&A...392..377B}.

The distribution of galaxies scatters around the normal SFGs values with some outliers. Among the outliers we found ultra-steep spectrum galaxies (USS; $\alpha > 1$), typically with low $L_{\rm{radio}}$. A similar distribution of galaxies was found in \citet{2014MNRAS.442..577T}. The nature of USS galaxies is debated and beyond the scope of our work. Early-stage mergers are capable of steepening the radio spectrum enhancing the radio emission \citep{2013ApJ...777...58M}. \citet{2014MNRAS.442..577T} also argued that an alternative scenario of USS are galaxies with radio jet emission truncated by interactions with dense gas in their environments \citep{1998PASP..110..493O}.

The location of cSFGs seems to be slightly biased towards \citet{2002A&A...392..377B} tracks at older ages (220--400\,Myr). We checked whether there exists a trend in $\Delta \Sigma_{\rm{QGs}}$ and, thus, in compactness along \citet{2002A&A...392..377B} tracks. Following \citet{2014MNRAS.442..577T}, we divide the parameter space overlapping with \citet{2002A&A...392..377B} tracks in three boxes representing young (0--20\,Myr), middle-aged (33--130\,Myr), and old (220--400\,Myr) SBs. We considered all extended SFGs and cSFGs, removing the galaxies classified as AGN since \citet{2002A&A...392..377B} tracks refer only to pure star formation and the contribution from the AGN to the radio spectrum could bias the interpretation. We found that $\Delta \Sigma_{\rm{QGs}}$ grows with the age of the starburst episode (see Table~\ref{tab:radio}), growing from $\Delta \Sigma_{\rm{QGs}} = 0.137$ to $\Delta \Sigma_{\rm{QGs}} = 0.55 \pm 0.63$. These values correspond to the median and the uncertainty is given by the median absolute deviation (MAD). Note that since the increase in compactness is a continuous function, we expect the scatter of each bin given by the MAD to overlap (as in Figure~\ref{fig:dms_dst}). We did not find a similar trend in the case of the stellar mass as reported in \citet{2014MNRAS.442..577T}, although to make a proper comparison selection effects should be considered.

\begin{deluxetable}{lccc}
\tabletypesize{\scriptsize}
\tablecaption{Properties of Galaxies at Different Age Bins} \label{tab:radio}
\tablehead{\colhead{Age} & \colhead{$\log (M_{\rm{*}}/M_{\odot})$} & \colhead{$\Delta \rm{MS}$} & \colhead{$\Delta \Sigma_{\rm{QGs}}$} \\
\colhead{(Myr)} & \colhead{} & \colhead{} & \colhead{}}
\startdata
0--20 & 10.67 & 6.44 & 0.137 \\
33--130 & 10.62 $\pm$ 0.37 & 3.5 $\pm$ 1.1 & 0.21 $\pm$ 0.20 \\
220--400 & 10.66 $\pm$ 0.19 & 5.14 $\pm$ 0.69 & 0.55 $\pm$ 0.63 \\
\enddata
\tablenotetext{}{The uncertainties refer to the MAD of the galaxies in each bin and, thus, since in the young (0--20\,Myr) age bin there is just one galaxy no dispersion is shown. AGN are excluded.}
\end{deluxetable}

Overall, we find a trend of increasing compactness with the age evolution of a starburst episode, leading to cSFGs at the final stages. This indicates that cSFGs could be old SBs, while extended SFGs could be a mix of normal SFGs and young SBs.

\section{Discussion} \label{sec:discussion}

\subsection{Compactness, Star Formation, and the Diagnostics of Burstiness} \label{subsec:disc_diag}

\begin{figure}
\begin{center}
\includegraphics[width=\columnwidth]{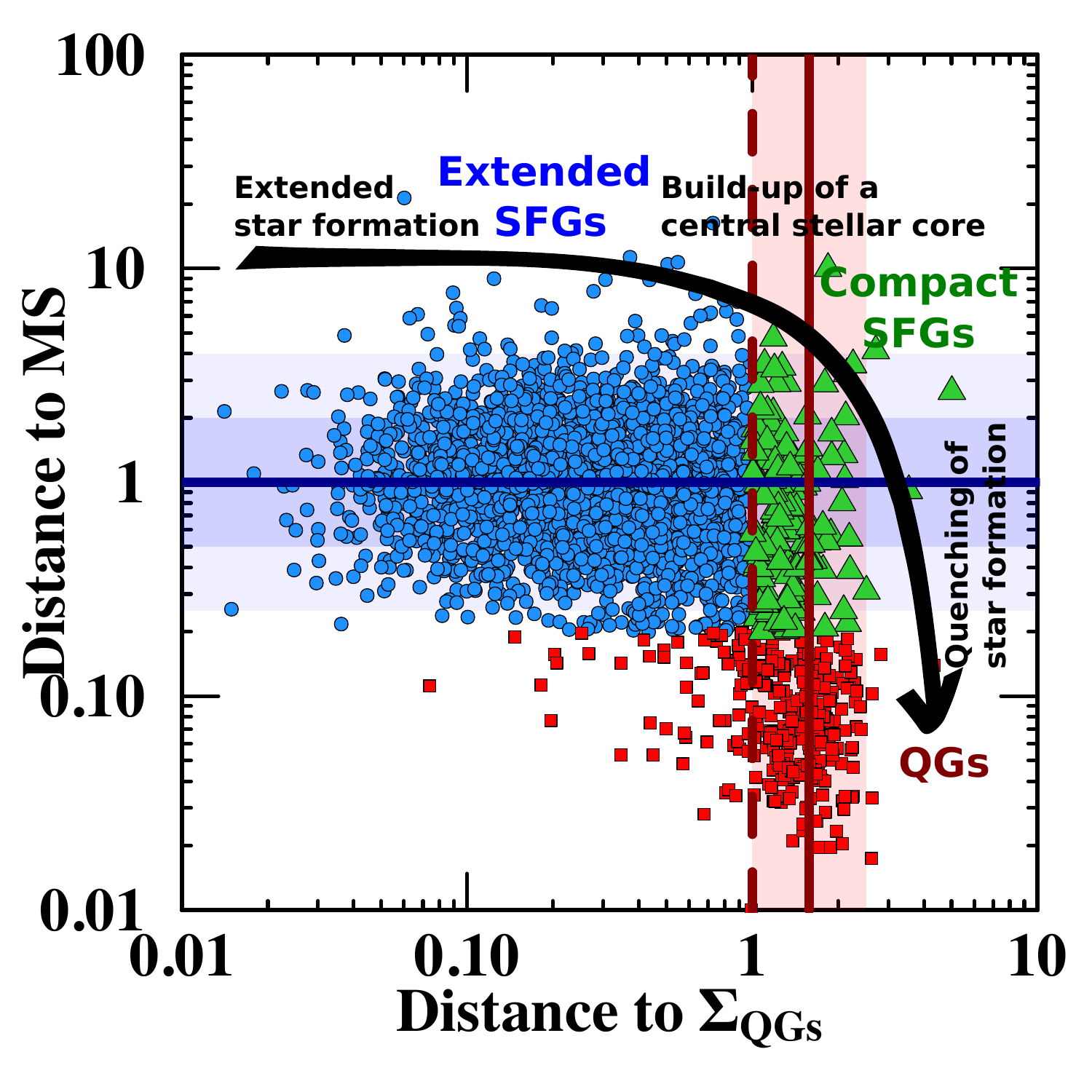}
\caption{Outline of the general evolutionary trend of the galaxy population.}
\label{fig:galevol_schema}
\end{center}
\end{figure}

In Section~\ref{sec:c_sf} we presented the distribution of the general galaxy population in the $\Delta \rm{MS}$-$\Delta \Sigma_{\rm{QGs}}$ plane. The general evolutionary trend of the galaxy population is given by the L-shape, previously reported in \citet{2017ApJ...840...47B}. Extended SFGs become compact before they quench (see Figure~\ref{fig:galevol_schema}, for a schematic plot). We stress that SFGs do not evolve into QGs at the same epoch (i.e., redshift), but into QGs at later times.

Furthermore, we showed two additional behaviours of the general evolutionary trend of the galaxy population: 1) galaxies start to transition smoothly towards quiescence, as the median sSFR decreases continuously for increasing compactness; 2) at least some of the SFGs galaxies become compact by increasing their sSFR and going above the scatter of the MS, given the large scatter and upper extreme values around the $\Delta \Sigma_{\rm{QGs}} \sim 1.0$ compactness transition threshold.

In Section~\ref{sec:csfgs_mssb} we examined three indicators of the burstiness of star formation: 1) SFE, 2) ISM, and 3) radio emission.

Regarding SFE there is no difference between cSFGs and extended SFGs. The similar values for cSFGs and extended SFGs extend to various regimes. There are both cSFGs and extended SFGs with low SFE due to high gas fractions, also both cSFGs and extended SFGs with enhanced SFE, some of which have enhanced SFE due to low gas fractions (see Figure~\ref{fig:sfe}). All together, it suggests that compactness could arise from different origins, like an extended normal SFG with low efficiency and a large gas reservoir that is secularly growing its stellar core, or an extended SB with enhanced efficiency that is rapidly consuming its gas reservoir growing its stellar core.

In terms of the ISM, the mild excitation and density values for extended SFGs in the upper-MS (see Figure~\ref{fig:ism} and Table~\ref{tab:ism}) are in line with \citet{2016MNRAS.457.2790T} scenario, which related the scatter of the MS to the evolution of galaxies following compaction events as part of the secular evolution of SFGs \citep[e.g.,][]{2013MNRAS.435..999D,2015MNRAS.450.2327Z,2016MNRAS.457.2790T}. If coming from the product of a rapid starburst-like event the latent ISM has similar properties to that of upper-MS normal SFGs.

On the other hand, regarding the radio emission diagnostic the increasing compactness with the expected age evolution of the radio emission in SBs leads to the conclusion of cSFGs could be old SBs. Note also that most of the galaxies are above the scatter of the MS in this part of the analysis, particularly in the old age bin. It suggests that the galaxies displaying high sSFR going above the scatter of the MS when becoming cSFGs (see Figure~\ref{fig:dms_dst}) are old SBs.

The analysis carried out in our work was performed in an unresolved fashion. This could be the reason of the apparent contradictory conclusions drawn from the SFE and ISM diagnostics versus the radio emission diagnostic. The conclusions drawn from the ISM and SFE diagnostics can be reconciled with that of the radio emission diagnostic if the SFE and ISM properties do not dominate the entire galaxy in an old SB phase. In that case the galaxy would not display an overall (unresolved) high SFE or SB-like ISM \citep[see][]{2019A&A...625A..65R}. This scenario would be also supported by the handful of resolved follow-up studies of the ISM of cSFGs, which indicate an undergoing nuclear starburst \citep[e.g.,][]{2016ApJ...832...19S,2017ApJ...851L..40B,2017A&A...602A..11P,2017ApJ...841L..25T,2018MNRAS.476.3956T}.

\subsection{Star Formation Efficiency and Selection Limits} \label{subsec:disc_sl}

\begin{figure}
\begin{center}
\includegraphics[width=\columnwidth]{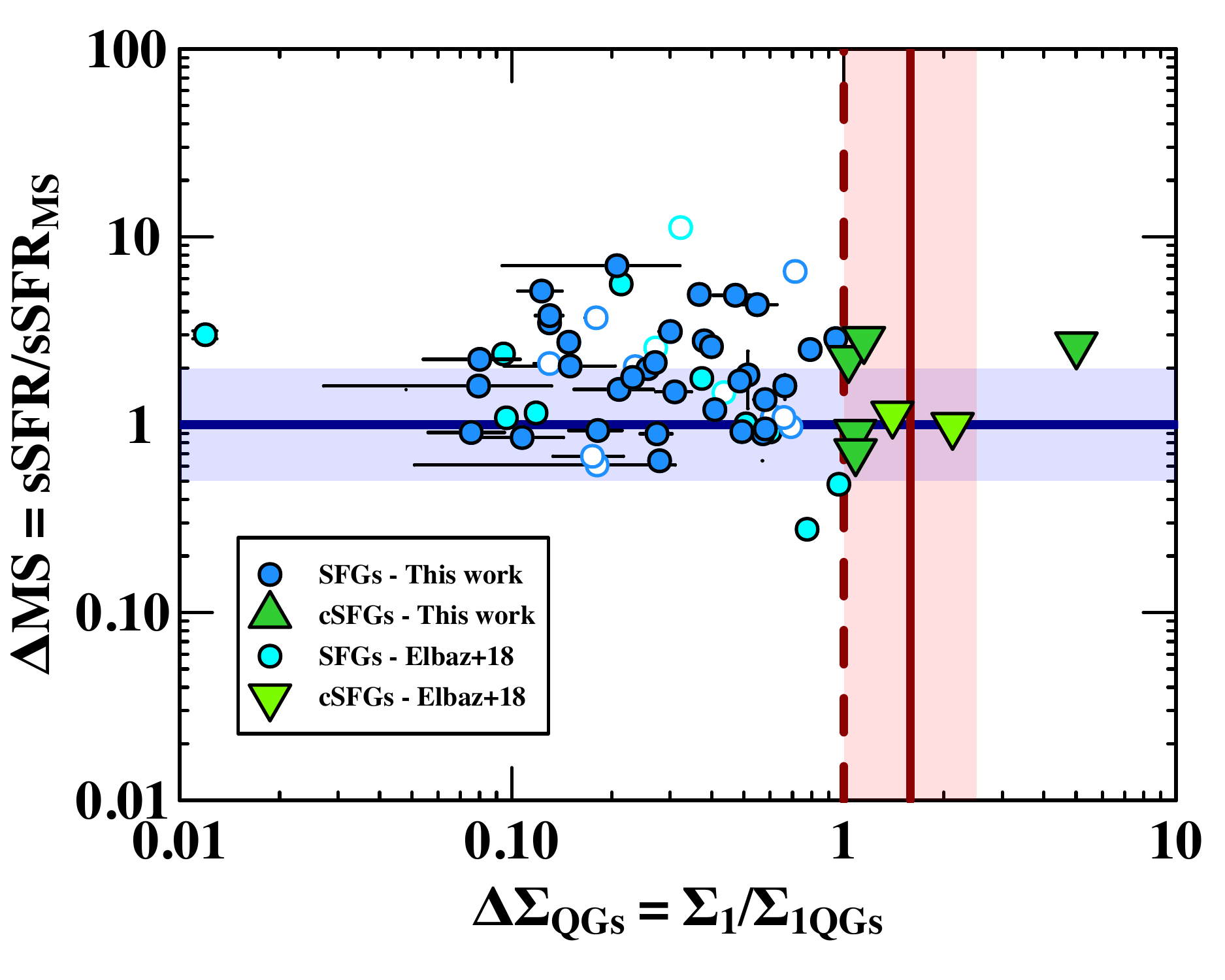}
\caption{$\Delta \rm{MS}$-$\Delta \Sigma_{\rm{QGs}}$ plane as presented in Figure~\ref{fig:dms_dst} for the R-J subset of the FIR sample and \citet{2018A&A...616A.110E} sample. The $\Delta \Sigma_{\rm{QGs}}$ uncertainties were calculated over 1000 realizations varying the structural parameters within their uncertainties (note that in some cases the uncertainties are smaller than the symbol size).}
\label{fig:dms_dst_sfe}
\end{center}
\end{figure}

We indicated in Section~\ref{subsec:sfe} that the combination of our sample with \citet{2018A&A...616A.110E} sample implied no difference between cSFGs and extended SFGs in terms of their SFE. However, it could be the case that \citet{2018A&A...616A.110E} galaxies are closer to the $\Delta \Sigma_{\rm{QGs}} \sim 1.0$ transition threshold, particularly for the subset with high SFE and low $f_{\rm{gas}}$ that could be on the last stage before becoming quiescent. In Figure~\ref{fig:dms_dst_sfe} we show the location of both samples in the $\Delta \rm{MS}$-$\Delta \Sigma_{\rm{QGs}}$ plane to explore this scenario. We did not find evidence that that the latter is the case as we do not see any distinction between our sample and \citet{2018A&A...616A.110E} sample in the $\Delta \rm{MS}$-$\Delta \Sigma_{\rm{QGs}}$ plane.

In Section~\ref{subsec:sfe} we also mentioned that our sample and \citet{2018A&A...616A.110E} sample are located in distinct complementary regions in the different diagrams shown Figure~\ref{fig:sfe}. For instance, we did not find galaxies with simultaneously high SFE and low $f_{\rm{gas}}$ within our sample, the galaxies described in \citet{2018A&A...616A.110E} as SBs that also fall within the scatter of the MS. \citet{2018A&A...616A.110E} galaxies occupy a parameter space offset form our sample, which made us consider the possibility that our selection is biased against the detection of SBs within the MS.

\begin{figure*}
\begin{center}
\includegraphics[width=\textwidth]{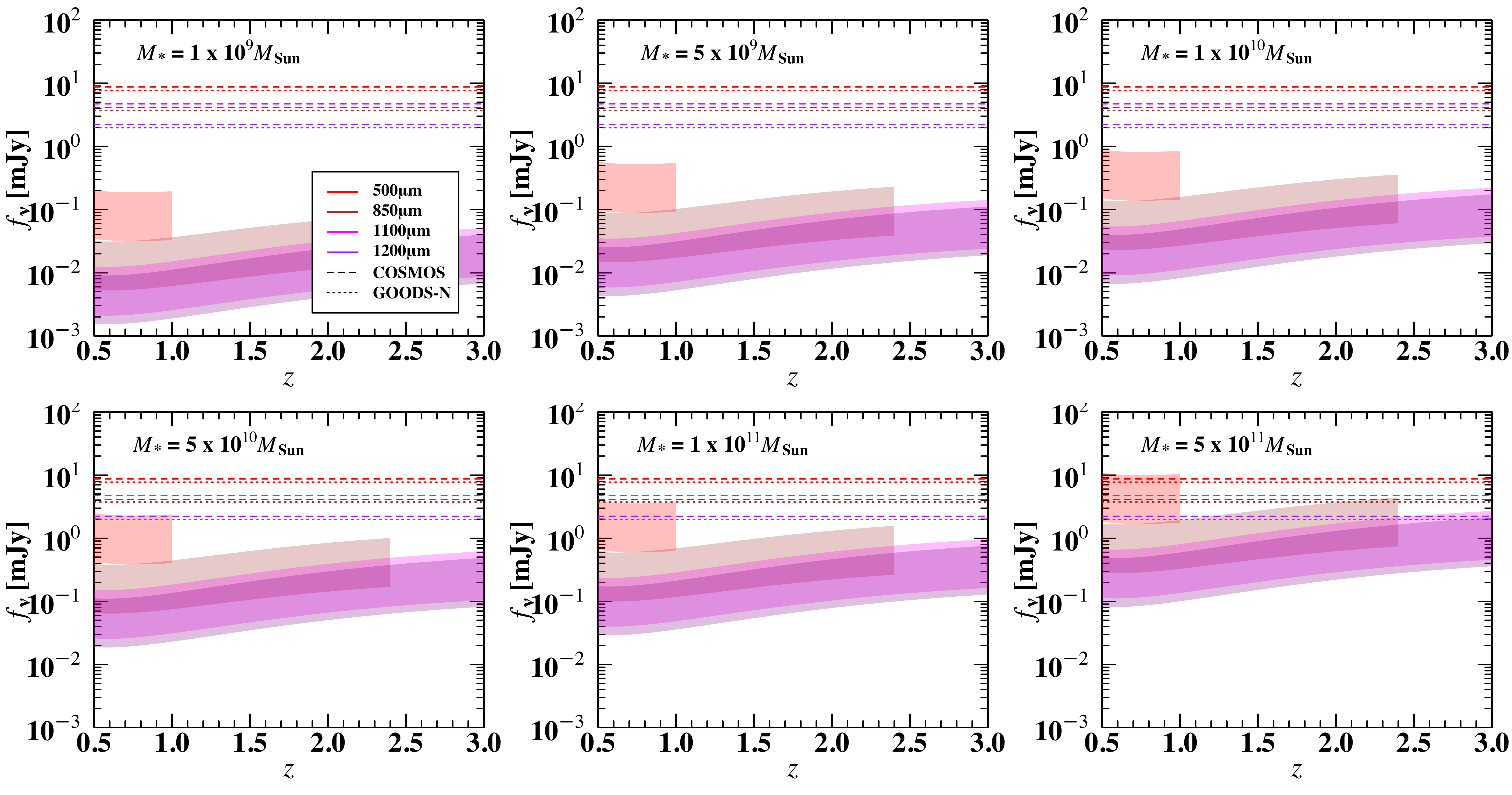}
\caption{Expected flux densities as a function of redshift for a galaxy with SFE three times higher ($M_{\rm{gas}}$ three times lower) than the average normal SFGs SFE trend for a fixed stellar mass. The width of the shaded areas represent the SFR scatter equivalent to $\Delta \rm{MS} = 3$ above and below the MS. The different bands plotted are those available in the COSMOS and GOODS-North fields at $\lambda_{\rm{rest}} > 250$\,$\mu$m for our redshift limits. The horizontal lines represent the 3$\sigma$ detection limits for the different bands (color-coded as the legend shows) for COSMOS (dashed lines) and GOODS-North (dotted lines). Note that in the case of GOODS-North the 1100\,mm and 1200\,mm bands are combined and, thus, the detection limit refers to 1160\,mm.}
\label{fig:sel_limits}
\end{center}
\end{figure*}

The R-J subset of the FIR sample was selected to fulfill a detection criteria in the R-J side of the FIR SED; therefore, establishing a detection limit for the different bands available in the R-J side of the "super-deblended" FIR catalogs in COSMOS and GOODS-North. For $0.5 < z < 3.0$ galaxies these bands are \textit{Herschel}/SPIRE 500\,$\mu$m, SCUBA 850\,$\mu$m, AzTEC 1.1\,mm, and MAMBO 1.2\,mm. We explored the required fluxes in these bands as a function of redshift to be able to detect galaxies that, while located within the scatter of the MS, exhibited enhanced SFE. We employed \citet{2016ApJ...820...83S} technique to predict the single band flux measurement of the dust continuum expected for a given $M_{\rm{gas}}$. In Figure~\ref{fig:sel_limits} we plot the predicted fluxes for the different bands as function of redshift for galaxies with a SFE three times higher ($M_{\rm{gas}}$ three times lower) than the normal SFGs SFE trend at fixed stellar mass for a range of SFRs equivalent to $\Delta \rm{MS} = 3$ both above and below the MS. Only galaxies with $M_{\rm{*}} \geq 5 \times 10^{11}$\,$M_{\odot}$ start to be detectable. These detection limits are consistent with the fact that we did not detect galaxies with enhanced SFE within the scatter of the MS. In Figure~\ref{fig:sfe} we show as a reference the most favorable $f_{\rm{gas}}$ limit (gray shaded region), which would correspond to a galaxy that has a stellar mass as high as the highest stellar mass of the sample ($\log M_{\rm{*}} = 11.73$) and located at a redshift such that the predicted detectable $f_{\rm{gas}}$ given the flux limits at the different bands in the two fields results in a minimum. Even in this extreme case, since there is not a galaxy in our sample that fulfills all the criteria at the same time, justifies that we missed galaxies with simultaneously high SFE and low $f_{\rm{gas}}$, like the ones presented in \citet{2018A&A...616A.110E}.

Therefore, we did not find SB-like SFE within the MS due to the detection limits in the catalogs.

\section{Summary and Conclusions} \label{sec:summary}

In this work we studied the general population of galaxies based on their location with respect to fundamental star-forming and structural relations, and classified them in extended, compact SFGs, and QGs. Based on a methodology of three diagnostics of the burstiness of star formation: 1) SFE, 2) ISM, and 3) radio emission, we aimed at studying whether cSFGs can be considered normal SFGs or SBs. As a proposed immediate transition population towards QGs, unveiling the nature of cSFGs implies understanding how the build-up of compact stellar cores and subsequent quenching of star formation happens. If cSFGs were normal SFGs it would point towards a secular transition towards quiescence and, conversely, a SB nature of cSFGs would point towards a more rapid transition towards quiescence. In summary we found:

\begin{itemize}

\item The distribution of galaxies in the $\Delta \rm{MS}$-$\Delta \Sigma_{\rm{QGs}}$ plane reveal that galaxies transition smoothly towards quiescence with increasing compactness. Some extended SFGs quench forming extended QGs. Most of the extended SFGs become compact before they quench, in agreement with \citep{2017ApJ...840...47B}. Furthermore, at least some galaxies become compact going above the scatter of the MS.

\item The MS is dominated by extended SFGs. However, SFGs with increasing compactness that are transitioning to quiescence contribute to lower the normalization of the MS.

\item There is no evidence for a distinct SFE in cSFGs and extended SFGs, suggesting that both secular and rapid evolution processes could generate cSFGs.

\item Extended SFGs located slightly above the MS (upper-MS galaxies) have ISM properties (CO excitation, density of the neutral gas, and strength of the ultraviolet) similar to lower envelope of SB-like ISM properties, and seem also similar to those of cSFG (with the caveat of the small sample sizes and the lack of ISM characterization for cSFGs). This suggest that the growth of a compact stellar core leading to cSFGs could happen secularly. Another explanation could be that if coming from a rapid starburst event, the latent ISM in cSFGs retains similar properties to that of upper-MS normal SFGs.

\item There is evidence for a trend in increasing compactness with the expected age evolution of the radio emission in SBs, indicating that cSFGs could be old SBs, while extended SFGs could be a mix of normal SFGs and young SBs.

The relative importance of a secular or rapid transition towards quiescence as a function of redshift remains to be completely understood. The apparent contradictory conclusions drawn from the SFE and ISM diagnostics versus the radio emission diagnostic can be reconciled if the SFE and ISM properties do not dominate the entire galaxy in an old SB phase, in agreement with resolved follow-up studies in the literature. We suggest that cSFGs could be SBs winding down and eventually crossing the main sequence towards quiescence.

\end{itemize}

\acknowledgments

We are very greatful to D. Elbaz, N. M. F\"orster Schreiber, and D. Watson for the assessment of this work. We thank A. Bressan and A. P. Thomson for their support on the starburst evolutionary tracks in radio. We are grateful to the anonymous referee, whose comments have been very useful to improving our work.

C.G.G and S.T. acknowledge support from the European Research Council (ERC) Consolidator Grant funding scheme (project ConTExt, grant number: 648179). G.M. and F.V. acknowledge the Villum Fonden research grant 13160 "Gas to stars, stars to dust: tracing star formation across cosmic time". A.M. is supported by the Dunlap Fellowship through an endowment established by the David Dunlap family and the University of Toronto.

The Cosmic Dawn Center is funded by the Danish National Research Foundation.

\bibliographystyle{aasjournal}{}
\bibliography{csfgs.bib}{}

\begin{thebibliography}{}
\expandafter\ifx\csname natexlab\endcsname\relax\def\natexlab#1{#1}\fi
\providecommand{\url}[1]{\href{#1}{#1}}
\providecommand{\dodoi}[1]{doi:~\href{http://doi.org/#1}{\nolinkurl{#1}}}
\providecommand{\doeprint}[1]{\href{http://ascl.net/#1}{\nolinkurl{http://ascl.net/#1}}}
\providecommand{\doarXiv}[1]{\href{https://arxiv.org/abs/#1}{\nolinkurl{https://arxiv.org/abs/#1}}}

\bibitem[{{Alaghband-Zadeh} {et~al.}(2013){Alaghband-Zadeh}, {Chapman},
  {Swinbank}, {Smail}, {Danielson}, {Decarli}, {Ivison}, {Meijerink}, {Weiss},
  \& {van der Werf}}]{2013MNRAS.435.1493A}
{Alaghband-Zadeh}, S., {Chapman}, S.~C., {Swinbank}, A.~M., {et~al.} 2013,
  \mnras, 435, 1493, \dodoi{10.1093/mnras/stt1390}

\bibitem[{{Avni}(1976)}]{1976ApJ...210..642A}
{Avni}, Y. 1976, \apj, 210, 642, \dodoi{10.1086/154870}

\bibitem[{{Barro} {et~al.}(2013){Barro}, {Faber}, {P{\'e}rez-Gonz{\'a}lez},
  {Koo}, {Williams}, {Kocevski}, {Trump}, {Mozena}, {McGrath}, {van der Wel},
  {Wuyts}, {Bell}, {Croton}, {Ceverino}, {Dekel}, {Ashby}, {Cheung},
  {Ferguson}, {Fontana}, {Fang}, {Giavalisco}, {Grogin}, {Guo}, {Hathi},
  {Hopkins}, {Huang}, {Koekemoer}, {Kartaltepe}, {Lee}, {Newman}, {Porter},
  {Primack}, {Ryan}, {Rosario}, {Somerville}, {Salvato}, \&
  {Hsu}}]{2013ApJ...765..104B}
{Barro}, G., {Faber}, S.~M., {P{\'e}rez-Gonz{\'a}lez}, P.~G., {et~al.} 2013,
  \apj, 765, 104, \dodoi{10.1088/0004-637X/765/2/104}

\bibitem[{{Barro} {et~al.}(2014){Barro}, {Faber}, {P{\'e}rez-Gonz{\'a}lez},
  {Pacifici}, {Trump}, {Koo}, {Wuyts}, {Guo}, {Bell}, {Dekel}, {Porter},
  {Primack}, {Ferguson}, {Ashby}, {Caputi}, {Ceverino}, {Croton}, {Fazio},
  {Giavalisco}, {Hsu}, {Kocevski}, {Koekemoer}, {Kurczynski}, {Kollipara},
  {Lee}, {McIntosh}, {McGrath}, {Moody}, {Somerville}, {Papovich}, {Salvato},
  {Santini}, {Tal}, {van der Wel}, {Williams}, {Willner}, \&
  {Zolotov}}]{2014ApJ...791...52B}
---. 2014, \apj, 791, 52, \dodoi{10.1088/0004-637X/791/1/52}

\bibitem[{{Barro} {et~al.}(2017{\natexlab{a}}){Barro}, {Faber}, {Koo}, {Dekel},
  {Fang}, {Trump}, {P{\'e}rez-Gonz{\'a}lez}, {Pacifici}, {Primack},
  {Somerville}, {Yan}, {Guo}, {Liu}, {Ceverino}, {Kocevski}, \&
  {McGrath}}]{2017ApJ...840...47B}
{Barro}, G., {Faber}, S.~M., {Koo}, D.~C., {et~al.} 2017{\natexlab{a}}, \apj,
  840, 47, \dodoi{10.3847/1538-4357/aa6b05}

\bibitem[{{Barro} {et~al.}(2017{\natexlab{b}}){Barro}, {Kriek},
  {P{\'e}rez-Gonz{\'a}lez}, {Diaz-Santos}, {Price}, {Rujopakarn}, {Pandya},
  {Koo}, {Faber}, {Dekel}, {Primack}, \& {Kocevski}}]{2017ApJ...851L..40B}
{Barro}, G., {Kriek}, M., {P{\'e}rez-Gonz{\'a}lez}, P.~G., {et~al.}
  2017{\natexlab{b}}, \apjl, 851, L40, \dodoi{10.3847/2041-8213/aa9f0d}

\bibitem[{{Berta} {et~al.}(2016){Berta}, {Lutz}, {Genzel},
  {F{\"o}rster-Schreiber}, \& {Tacconi}}]{2016A&A...587A..73B}
{Berta}, S., {Lutz}, D., {Genzel}, R., {F{\"o}rster-Schreiber}, N.~M., \&
  {Tacconi}, L.~J. 2016, \aap, 587, A73, \dodoi{10.1051/0004-6361/201527746}

\bibitem[{{Bothwell} {et~al.}(2013){Bothwell}, {Smail}, {Chapman}, {Genzel},
  {Ivison}, {Tacconi}, {Alaghband-Zadeh}, {Bertoldi}, {Blain}, {Casey}, {Cox},
  {Greve}, {Lutz}, {Neri}, {Omont}, \& {Swinbank}}]{2013MNRAS.429.3047B}
{Bothwell}, M.~S., {Smail}, I., {Chapman}, S.~C., {et~al.} 2013, \mnras, 429,
  3047, \dodoi{10.1093/mnras/sts562}

\bibitem[{{Brammer} {et~al.}(2012){Brammer}, {van Dokkum}, {Franx},
  {Fumagalli}, {Patel}, {Rix}, {Skelton}, {Kriek}, {Nelson}, {Schmidt},
  {Bezanson}, {da Cunha}, {Erb}, {Fan}, {F{\"o}rster Schreiber}, {Illingworth},
  {Labb{\'e}}, {Leja}, {Lundgren}, {Magee}, {Marchesini}, {McCarthy},
  {Momcheva}, {Muzzin}, {Quadri}, {Steidel}, {Tal}, {Wake}, {Whitaker}, \&
  {Williams}}]{2012ApJS..200...13B}
{Brammer}, G.~B., {van Dokkum}, P.~G., {Franx}, M., {et~al.} 2012, \apjs, 200,
  13, \dodoi{10.1088/0067-0049/200/2/13}

\bibitem[{{Bressan} {et~al.}(2002){Bressan}, {Silva}, \&
  {Granato}}]{2002A&A...392..377B}
{Bressan}, A., {Silva}, L., \& {Granato}, G.~L. 2002, \aap, 392, 377,
  \dodoi{10.1051/0004-6361:20020960}

\bibitem[{{Brinchmann} {et~al.}(2004){Brinchmann}, {Charlot}, {White},
  {Tremonti}, {Kauffmann}, {Heckman}, \& {Brinkmann}}]{2004MNRAS.351.1151B}
{Brinchmann}, J., {Charlot}, S., {White}, S.~D.~M., {et~al.} 2004, \mnras, 351,
  1151, \dodoi{10.1111/j.1365-2966.2004.07881.x}

\bibitem[{{Chabrier}(2003)}]{2003PASP..115..763C}
{Chabrier}, G. 2003, \pasp, 115, 763, \dodoi{10.1086/376392}

\bibitem[{{Cimatti} {et~al.}(2008){Cimatti}, {Cassata}, {Pozzetti}, {Kurk},
  {Mignoli}, {Renzini}, {Daddi}, {Bolzonella}, {Brusa}, {Rodighiero},
  {Dickinson}, {Franceschini}, {Zamorani}, {Berta}, {Rosati}, \&
  {Halliday}}]{2008A&A...482...21C}
{Cimatti}, A., {Cassata}, P., {Pozzetti}, L., {et~al.} 2008, \aap, 482, 21,
  \dodoi{10.1051/0004-6361:20078739}

\bibitem[{{Civano} {et~al.}(2016){Civano}, {Marchesi}, {Comastri}, {Urry},
  {Elvis}, {Cappelluti}, {Puccetti}, {Brusa}, {Zamorani}, {Hasinger},
  {Aldcroft}, {Alexander}, {Allevato}, {Brunner}, {Capak}, {Finoguenov},
  {Fiore}, {Fruscione}, {Gilli}, {Glotfelty}, {Griffiths}, {Hao}, {Harrison},
  {Jahnke}, {Kartaltepe}, {Karim}, {LaMassa}, {Lanzuisi}, {Miyaji}, {Ranalli},
  {Salvato}, {Sargent}, {Scoville}, {Schawinski}, {Schinnerer}, {Silverman},
  {Smolcic}, {Stern}, {Toft}, {Trakhtenbrot}, {Treister}, \&
  {Vignali}}]{2016ApJ...819...62C}
{Civano}, F., {Marchesi}, S., {Comastri}, A., {et~al.} 2016, \apj, 819, 62,
  \dodoi{10.3847/0004-637X/819/1/62}

\bibitem[{{Condon}(1992)}]{1992ARA&A..30..575C}
{Condon}, J.~J. 1992, \araa, 30, 575,
  \dodoi{10.1146/annurev.aa.30.090192.003043}

\bibitem[{{Daddi} {et~al.}(2007){Daddi}, {Dickinson}, {Morrison}, {Chary},
  {Cimatti}, {Elbaz}, {Frayer}, {Renzini}, {Pope}, {Alexander}, {Bauer},
  {Giavalisco}, {Huynh}, {Kurk}, \& {Mignoli}}]{2007ApJ...670..156D}
{Daddi}, E., {Dickinson}, M., {Morrison}, G., {et~al.} 2007, \apj, 670, 156,
  \dodoi{10.1086/521818}

\bibitem[{{Daddi} {et~al.}(2010{\natexlab{a}}){Daddi}, {Bournaud}, {Walter},
  {Dannerbauer}, {Carilli}, {Dickinson}, {Elbaz}, {Morrison}, {Riechers},
  {Onodera}, {Salmi}, {Krips}, \& {Stern}}]{2010ApJ...713..686D}
{Daddi}, E., {Bournaud}, F., {Walter}, F., {et~al.} 2010{\natexlab{a}}, \apj,
  713, 686, \dodoi{10.1088/0004-637X/713/1/686}

\bibitem[{{Daddi} {et~al.}(2010{\natexlab{b}}){Daddi}, {Elbaz}, {Walter},
  {Bournaud}, {Salmi}, {Carilli}, {Dannerbauer}, {Dickinson}, {Monaco}, \&
  {Riechers}}]{2010ApJ...714L.118D}
{Daddi}, E., {Elbaz}, D., {Walter}, F., {et~al.} 2010{\natexlab{b}}, \apjl,
  714, L118, \dodoi{10.1088/2041-8205/714/1/L118}

\bibitem[{{Daddi} {et~al.}(2015){Daddi}, {Dannerbauer}, {Liu}, {Aravena},
  {Bournaud}, {Walter}, {Riechers}, {Magdis}, {Sargent}, {B{\'e}thermin},
  {Carilli}, {Cibinel}, {Dickinson}, {Elbaz}, {Gao}, {Gobat}, {Hodge}, \&
  {Krips}}]{2015A&A...577A..46D}
{Daddi}, E., {Dannerbauer}, H., {Liu}, D., {et~al.} 2015, \aap, 577, A46,
  \dodoi{10.1051/0004-6361/201425043}

\bibitem[{{Davies} {et~al.}(2003){Davies}, {Sternberg}, {Lehnert}, \&
  {Tacconi-Garman}}]{2003ApJ...597..907D}
{Davies}, R.~I., {Sternberg}, A., {Lehnert}, M., \& {Tacconi-Garman}, L.~E.
  2003, \apj, 597, 907, \dodoi{10.1086/378634}

\bibitem[{{de Jong} {et~al.}(1985){de Jong}, {Klein}, {Wielebinski}, \&
  {Wunderlich}}]{1985A&A...147L...6D}
{de Jong}, T., {Klein}, U., {Wielebinski}, R., \& {Wunderlich}, E. 1985, \aap,
  147, L6

\bibitem[{{Dekel} {et~al.}(2013){Dekel}, {Zolotov}, {Tweed}, {Cacciato},
  {Ceverino}, \& {Primack}}]{2013MNRAS.435..999D}
{Dekel}, A., {Zolotov}, A., {Tweed}, D., {et~al.} 2013, \mnras, 435, 999,
  \dodoi{10.1093/mnras/stt1338}

\bibitem[{{Del Moro} {et~al.}(2013){Del Moro}, {Alexander}, {Mullaney},
  {Daddi}, {Pannella}, {Bauer}, {Pope}, {Dickinson}, {Elbaz}, {Barthel},
  {Garrett}, {Brandt}, {Charmandaris}, {Chary}, {Dasyra}, {Gilli}, {Hickox},
  {Hwang}, {Ivison}, {Juneau}, {Le Floc'h}, {Luo}, {Morrison}, {Rovilos},
  {Sargent}, \& {Xue}}]{2013A&A...549A..59D}
{Del Moro}, A., {Alexander}, D.~M., {Mullaney}, J.~R., {et~al.} 2013, \aap,
  549, A59, \dodoi{10.1051/0004-6361/201219880}

\bibitem[{{Di Matteo} {et~al.}(2008){Di Matteo}, {Bournaud}, {Martig},
  {Combes}, {Melchior}, \& {Semelin}}]{2008A&A...492...31D}
{Di Matteo}, P., {Bournaud}, F., {Martig}, M., {et~al.} 2008, \aap, 492, 31,
  \dodoi{10.1051/0004-6361:200809480}

\bibitem[{{Dickinson} {et~al.}(2003){Dickinson}, {Giavalisco}, \& {GOODS
  Team}}]{2003mglh.conf..324D}
{Dickinson}, M., {Giavalisco}, M., \& {GOODS Team}. 2003, in The Mass of
  Galaxies at Low and High Redshift, ed. R.~{Bender} \& A.~{Renzini}, 324

\bibitem[{{Draine} \& {Li}(2007)}]{2007ApJ...657..810D}
{Draine}, B.~T., \& {Li}, A. 2007, \apj, 657, 810, \dodoi{10.1086/511055}

\bibitem[{{Elbaz} {et~al.}(2007){Elbaz}, {Daddi}, {Le Borgne}, {Dickinson},
  {Alexander}, {Chary}, {Starck}, {Brandt}, {Kitzbichler}, {MacDonald},
  {Nonino}, {Popesso}, {Stern}, \& {Vanzella}}]{2007A&A...468...33E}
{Elbaz}, D., {Daddi}, E., {Le Borgne}, D., {et~al.} 2007, \aap, 468, 33,
  \dodoi{10.1051/0004-6361:20077525}

\bibitem[{{Elbaz} {et~al.}(2018){Elbaz}, {Leiton}, {Nagar}, {Okumura},
  {Franco}, {Schreiber}, {Pannella}, {Wang}, {Dickinson}, {D{\'{\i}}az-Santos},
  {Ciesla}, {Daddi}, {Bournaud}, {Magdis}, {Zhou}, \&
  {Rujopakarn}}]{2018A&A...616A.110E}
{Elbaz}, D., {Leiton}, R., {Nagar}, N., {et~al.} 2018, \aap, 616, A110,
  \dodoi{10.1051/0004-6361/201732370}

\bibitem[{{Feldmann} \& {Mayer}(2015)}]{2015MNRAS.446.1939F}
{Feldmann}, R., \& {Mayer}, L. 2015, \mnras, 446, 1939,
  \dodoi{10.1093/mnras/stu2207}

\bibitem[{{Fu} {et~al.}(2013){Fu}, {Cooray}, {Feruglio}, {Ivison}, {Riechers},
  {Gurwell}, {Bussmann}, {Harris}, {Altieri}, {Aussel}, {Baker}, {Bock},
  {Boylan-Kolchin}, {Bridge}, {Calanog}, {Casey}, {Cava}, {Chapman},
  {Clements}, {Conley}, {Cox}, {Farrah}, {Frayer}, {Hopwood}, {Jia}, {Magdis},
  {Marsden}, {Mart{\'{\i}}nez-Navajas}, {Negrello}, {Neri}, {Oliver}, {Omont},
  {Page}, {P{\'e}rez-Fournon}, {Schulz}, {Scott}, {Smith}, {Vaccari},
  {Valtchanov}, {Vieira}, {Viero}, {Wang}, {Wardlow}, \&
  {Zemcov}}]{2013Natur.498..338F}
{Fu}, H., {Cooray}, A., {Feruglio}, C., {et~al.} 2013, \nat, 498, 338,
  \dodoi{10.1038/nature12184}

\bibitem[{{Genzel} {et~al.}(2010){Genzel}, {Tacconi}, {Gracia-Carpio},
  {Sternberg}, {Cooper}, {Shapiro}, {Bolatto}, {Bouch{\'e}}, {Bournaud},
  {Burkert}, {Combes}, {Comerford}, {Cox}, {Davis}, {Schreiber},
  {Garcia-Burillo}, {Lutz}, {Naab}, {Neri}, {Omont}, {Shapley}, \&
  {Weiner}}]{2010MNRAS.407.2091G}
{Genzel}, R., {Tacconi}, L.~J., {Gracia-Carpio}, J., {et~al.} 2010, \mnras,
  407, 2091, \dodoi{10.1111/j.1365-2966.2010.16969.x}

\bibitem[{{G{\'o}mez-Guijarro} {et~al.}(2018){G{\'o}mez-Guijarro}, {Toft},
  {Karim}, {Magnelli}, {Magdis}, {Jim{\'e}nez-Andrade}, {Capak}, {Fraternali},
  {Fujimoto}, {Riechers}, {Schinnerer}, {Smol{\v c}i{\'c}}, {Aravena},
  {Bertoldi}, {Cortzen}, {Hasinger}, {Hu}, {Jones}, {Koekemoer}, {Lee},
  {McCracken}, {Micha{\l}owski}, {Navarrete}, {Povi{\'c}}, {Puglisi},
  {Romano-D{\'{\i}}az}, {Sheth}, {Silverman}, {Staguhn}, {Steinhardt},
  {Stockmann}, {Tanaka}, {Valentino}, {van Kampen}, \&
  {Zirm}}]{2018ApJ...856..121G}
{G{\'o}mez-Guijarro}, C., {Toft}, S., {Karim}, A., {et~al.} 2018, \apj, 856,
  121, \dodoi{10.3847/1538-4357/aab206}

\bibitem[{{Grogin} {et~al.}(2011){Grogin}, {Kocevski}, {Faber}, {Ferguson},
  {Koekemoer}, {Riess}, {Acquaviva}, {Alexander}, {Almaini}, {Ashby}, {Barden},
  {Bell}, {Bournaud}, {Brown}, {Caputi}, {Casertano}, {Cassata}, {Castellano},
  {Challis}, {Chary}, {Cheung}, {Cirasuolo}, {Conselice}, {Roshan Cooray},
  {Croton}, {Daddi}, {Dahlen}, {Dav{\'e}}, {de Mello}, {Dekel}, {Dickinson},
  {Dolch}, {Donley}, {Dunlop}, {Dutton}, {Elbaz}, {Fazio}, {Filippenko},
  {Finkelstein}, {Fontana}, {Gardner}, {Garnavich}, {Gawiser}, {Giavalisco},
  {Grazian}, {Guo}, {Hathi}, {H{\"a}ussler}, {Hopkins}, {Huang}, {Huang},
  {Jha}, {Kartaltepe}, {Kirshner}, {Koo}, {Lai}, {Lee}, {Li}, {Lotz}, {Lucas},
  {Madau}, {McCarthy}, {McGrath}, {McIntosh}, {McLure}, {Mobasher},
  {Moustakas}, {Mozena}, {Nandra}, {Newman}, {Niemi}, {Noeske}, {Papovich},
  {Pentericci}, {Pope}, {Primack}, {Rajan}, {Ravindranath}, {Reddy}, {Renzini},
  {Rix}, {Robaina}, {Rodney}, {Rosario}, {Rosati}, {Salimbeni}, {Scarlata},
  {Siana}, {Simard}, {Smidt}, {Somerville}, {Spinrad}, {Straughn}, {Strolger},
  {Telford}, {Teplitz}, {Trump}, {van der Wel}, {Villforth}, {Wechsler},
  {Weiner}, {Wiklind}, {Wild}, {Wilson}, {Wuyts}, {Yan}, \&
  {Yun}}]{2011ApJS..197...35G}
{Grogin}, N.~A., {Kocevski}, D.~D., {Faber}, S.~M., {et~al.} 2011, \apjs, 197,
  35, \dodoi{10.1088/0067-0049/197/2/35}

\bibitem[{{Groves} {et~al.}(2015){Groves}, {Schinnerer}, {Leroy}, {Galametz},
  {Walter}, {Bolatto}, {Hunt}, {Dale}, {Calzetti}, {Croxall}, \&
  {Kennicutt}}]{2015ApJ...799...96G}
{Groves}, B.~A., {Schinnerer}, E., {Leroy}, A., {et~al.} 2015, \apj, 799, 96,
  \dodoi{10.1088/0004-637X/799/1/96}

\bibitem[{{Helou} {et~al.}(1985){Helou}, {Soifer}, \&
  {Rowan-Robinson}}]{1985ApJ...298L...7H}
{Helou}, G., {Soifer}, B.~T., \& {Rowan-Robinson}, M. 1985, \apjl, 298, L7,
  \dodoi{10.1086/184556}

\bibitem[{{Ibar} {et~al.}(2010){Ibar}, {Ivison}, {Best}, {Coppin}, {Pope},
  {Smail}, \& {Dunlop}}]{2010MNRAS.401L..53I}
{Ibar}, E., {Ivison}, R.~J., {Best}, P.~N., {et~al.} 2010, \mnras, 401, L53,
  \dodoi{10.1111/j.1745-3933.2009.00786.x}

\bibitem[{{Ibar} {et~al.}(2009){Ibar}, {Ivison}, {Biggs}, {Lal}, {Best}, \&
  {Green}}]{2009MNRAS.397..281I}
{Ibar}, E., {Ivison}, R.~J., {Biggs}, A.~D., {et~al.} 2009, \mnras, 397, 281,
  \dodoi{10.1111/j.1365-2966.2009.14866.x}

\bibitem[{{Ivison} {et~al.}(2013){Ivison}, {Swinbank}, {Smail}, {Harris},
  {Bussmann}, {Cooray}, {Cox}, {Fu}, {Kov{\'a}cs}, {Krips}, {Narayanan},
  {Negrello}, {Neri}, {Pe{\~n}arrubia}, {Richard}, {Riechers}, {Rowlands},
  {Staguhn}, {Targett}, {Amber}, {Baker}, {Bourne}, {Bertoldi}, {Bremer},
  {Calanog}, {Clements}, {Dannerbauer}, {Dariush}, {De Zotti}, {Dunne},
  {Eales}, {Farrah}, {Fleuren}, {Franceschini}, {Geach}, {George}, {Helly},
  {Hopwood}, {Ibar}, {Jarvis}, {Kneib}, {Maddox}, {Omont}, {Scott}, {Serjeant},
  {Smith}, {Thompson}, {Valiante}, {Valtchanov}, {Vieira}, \& {van der
  Werf}}]{2013ApJ...772..137I}
{Ivison}, R.~J., {Swinbank}, A.~M., {Smail}, I., {et~al.} 2013, \apj, 772, 137,
  \dodoi{10.1088/0004-637X/772/2/137}

\bibitem[{{Jim{\'e}nez-Andrade} {et~al.}(2019){Jim{\'e}nez-Andrade},
  {Magnelli}, {Karim}, {Zamorani}, {Bondi}, {Schinnerer}, {Sargent},
  {Romano-D{\'\i}az}, {Novak}, {Lang}, {Bertoldi}, {Vardoulaki}, {Toft},
  {Smol{\v{c}}i{\'c}}, {Harrington}, {Leslie}, {Delhaize}, {Liu}, {Karoumpis},
  {Kartaltepe}, \& {Koekemoer}}]{2019A&A...625A.114J}
{Jim{\'e}nez-Andrade}, E.~F., {Magnelli}, B., {Karim}, A., {et~al.} 2019, \aap,
  625, A114, \dodoi{10.1051/0004-6361/201935178}

\bibitem[{{Jin} {et~al.}(2018){Jin}, {Daddi}, {Liu}, {Smol{\v c}i{\'c}},
  {Schinnerer}, {Calabr{\`o}}, {Gu}, {Delhaize}, {Delvecchio}, {Gao},
  {Salvato}, {Puglisi}, {Dickinson}, {Bertoldi}, {Sargent}, {Novak}, {Magdis},
  {Aretxaga}, {Wilson}, \& {Capak}}]{2018ApJ...864...56J}
{Jin}, S., {Daddi}, E., {Liu}, D., {et~al.} 2018, \apj, 864, 56,
  \dodoi{10.3847/1538-4357/aad4af}

\bibitem[{{Kauffmann} {et~al.}(2003){Kauffmann}, {Heckman}, {Tremonti},
  {Brinchmann}, {Charlot}, {White}, {Ridgway}, {Brinkmann}, {Fukugita}, {Hall},
  {Ivezi{\'c}}, {Richards}, \& {Schneider}}]{2003MNRAS.346.1055K}
{Kauffmann}, G., {Heckman}, T.~M., {Tremonti}, C., {et~al.} 2003, \mnras, 346,
  1055, \dodoi{10.1111/j.1365-2966.2003.07154.x}

\bibitem[{{Kaufman} {et~al.}(2006){Kaufman}, {Wolfire}, \&
  {Hollenbach}}]{2006ApJ...644..283K}
{Kaufman}, M.~J., {Wolfire}, M.~G., \& {Hollenbach}, D.~J. 2006, \apj, 644,
  283, \dodoi{10.1086/503596}

\bibitem[{{Kaufman} {et~al.}(1999){Kaufman}, {Wolfire}, {Hollenbach}, \&
  {Luhman}}]{1999ApJ...527..795K}
{Kaufman}, M.~J., {Wolfire}, M.~G., {Hollenbach}, D.~J., \& {Luhman}, M.~L.
  1999, \apj, 527, 795, \dodoi{10.1086/308102}

\bibitem[{{Kennicutt}(1998)}]{1998ARA&A..36..189K}
{Kennicutt}, Jr., R.~C. 1998, \araa, 36, 189,
  \dodoi{10.1146/annurev.astro.36.1.189}

\bibitem[{{Koekemoer} {et~al.}(2011){Koekemoer}, {Faber}, {Ferguson}, {Grogin},
  {Kocevski}, {Koo}, {Lai}, {Lotz}, {Lucas}, {McGrath}, {Ogaz}, {Rajan},
  {Riess}, {Rodney}, {Strolger}, {Casertano}, {Castellano}, {Dahlen},
  {Dickinson}, {Dolch}, {Fontana}, {Giavalisco}, {Grazian}, {Guo}, {Hathi},
  {Huang}, {van der Wel}, {Yan}, {Acquaviva}, {Alexander}, {Almaini}, {Ashby},
  {Barden}, {Bell}, {Bournaud}, {Brown}, {Caputi}, {Cassata}, {Challis},
  {Chary}, {Cheung}, {Cirasuolo}, {Conselice}, {Roshan Cooray}, {Croton},
  {Daddi}, {Dav{\'e}}, {de Mello}, {de Ravel}, {Dekel}, {Donley}, {Dunlop},
  {Dutton}, {Elbaz}, {Fazio}, {Filippenko}, {Finkelstein}, {Frazer}, {Gardner},
  {Garnavich}, {Gawiser}, {Gruetzbauch}, {Hartley}, {H{\"a}ussler},
  {Herrington}, {Hopkins}, {Huang}, {Jha}, {Johnson}, {Kartaltepe},
  {Khostovan}, {Kirshner}, {Lani}, {Lee}, {Li}, {Madau}, {McCarthy},
  {McIntosh}, {McLure}, {McPartland}, {Mobasher}, {Moreira}, {Mortlock},
  {Moustakas}, {Mozena}, {Nandra}, {Newman}, {Nielsen}, {Niemi}, {Noeske},
  {Papovich}, {Pentericci}, {Pope}, {Primack}, {Ravindranath}, {Reddy},
  {Renzini}, {Rix}, {Robaina}, {Rosario}, {Rosati}, {Salimbeni}, {Scarlata},
  {Siana}, {Simard}, {Smidt}, {Snyder}, {Somerville}, {Spinrad}, {Straughn},
  {Telford}, {Teplitz}, {Trump}, {Vargas}, {Villforth}, {Wagner}, {Wandro},
  {Wechsler}, {Weiner}, {Wiklind}, {Wild}, {Wilson}, {Wuyts}, \&
  {Yun}}]{2011ApJS..197...36K}
{Koekemoer}, A.~M., {Faber}, S.~M., {Ferguson}, H.~C., {et~al.} 2011, \apjs,
  197, 36, \dodoi{10.1088/0067-0049/197/2/36}

\bibitem[{{Lang} {et~al.}(2014){Lang}, {Wuyts}, {Somerville}, {F{\"o}rster
  Schreiber}, {Genzel}, {Bell}, {Brammer}, {Dekel}, {Faber}, {Ferguson},
  {Grogin}, {Kocevski}, {Koekemoer}, {Lutz}, {McGrath}, {Momcheva}, {Nelson},
  {Primack}, {Rosario}, {Skelton}, {Tacconi}, {van Dokkum}, \&
  {Whitaker}}]{2014ApJ...788...11L}
{Lang}, P., {Wuyts}, S., {Somerville}, R.~S., {et~al.} 2014, \apj, 788, 11,
  \dodoi{10.1088/0004-637X/788/1/11}

\bibitem[{{Liu} {et~al.}(2018){Liu}, {Daddi}, {Dickinson}, {Owen}, {Pannella},
  {Sargent}, {B{\'e}thermin}, {Magdis}, {Gao}, {Shu}, {Wang}, {Jin}, \&
  {Inami}}]{2018ApJ...853..172L}
{Liu}, D., {Daddi}, E., {Dickinson}, M., {et~al.} 2018, \apj, 853, 172,
  \dodoi{10.3847/1538-4357/aaa600}

\bibitem[{{Magdis} {et~al.}(2012){Magdis}, {Daddi}, {B{\'e}thermin}, {Sargent},
  {Elbaz}, {Pannella}, {Dickinson}, {Dannerbauer}, {da Cunha}, {Walter},
  {Rigopoulou}, {Charmandaris}, {Hwang}, \& {Kartaltepe}}]{2012ApJ...760....6M}
{Magdis}, G.~E., {Daddi}, E., {B{\'e}thermin}, M., {et~al.} 2012, \apj, 760, 6,
  \dodoi{10.1088/0004-637X/760/1/6}

\bibitem[{{Magdis} {et~al.}(2017){Magdis}, {Rigopoulou}, {Daddi}, {Bethermin},
  {Feruglio}, {Sargent}, {Dannerbauer}, {Dickinson}, {Elbaz}, {Gomez Guijarro},
  {Huang}, {Toft}, \& {Valentino}}]{2017A&A...603A..93M}
{Magdis}, G.~E., {Rigopoulou}, D., {Daddi}, E., {et~al.} 2017, \aap, 603, A93,
  \dodoi{10.1051/0004-6361/201731037}

\bibitem[{{Magnelli} {et~al.}(2015){Magnelli}, {Ivison}, {Lutz}, {Valtchanov},
  {Farrah}, {Berta}, {Bertoldi}, {Bock}, {Cooray}, {Ibar}, {Karim}, {Le
  Floc'h}, {Nordon}, {Oliver}, {Page}, {Popesso}, {Pozzi}, {Rigopoulou},
  {Riguccini}, {Rodighiero}, {Rosario}, {Roseboom}, {Wang}, \&
  {Wuyts}}]{2015A&A...573A..45M}
{Magnelli}, B., {Ivison}, R.~J., {Lutz}, D., {et~al.} 2015, \aap, 573, A45,
  \dodoi{10.1051/0004-6361/201424937}

\bibitem[{{Malhotra} {et~al.}(2001){Malhotra}, {Kaufman}, {Hollenbach},
  {Helou}, {Rubin}, {Brauher}, {Dale}, {Lu}, {Lord}, {Stacey}, {Contursi},
  {Hunter}, \& {Dinerstein}}]{2001ApJ...561..766M}
{Malhotra}, S., {Kaufman}, M.~J., {Hollenbach}, D., {et~al.} 2001, \apj, 561,
  766, \dodoi{10.1086/323046}

\bibitem[{{Marchesi} {et~al.}(2016){Marchesi}, {Civano}, {Elvis}, {Salvato},
  {Brusa}, {Comastri}, {Gilli}, {Hasinger}, {Lanzuisi}, {Miyaji}, {Treister},
  {Urry}, {Vignali}, {Zamorani}, {Allevato}, {Cappelluti}, {Cardamone},
  {Finoguenov}, {Griffiths}, {Karim}, {Laigle}, {LaMassa}, {Jahnke}, {Ranalli},
  {Schawinski}, {Schinnerer}, {Silverman}, {Smolcic}, {Suh}, \&
  {Trakhtenbrot}}]{2016ApJ...817...34M}
{Marchesi}, S., {Civano}, F., {Elvis}, M., {et~al.} 2016, \apj, 817, 34,
  \dodoi{10.3847/0004-637X/817/1/34}

\bibitem[{{Momcheva} {et~al.}(2016){Momcheva}, {Brammer}, {van Dokkum},
  {Skelton}, {Whitaker}, {Nelson}, {Fumagalli}, {Maseda}, {Leja}, {Franx},
  {Rix}, {Bezanson}, {Da Cunha}, {Dickey}, {F{\"o}rster Schreiber},
  {Illingworth}, {Kriek}, {Labb{\'e}}, {Ulf Lange}, {Lundgren}, {Magee},
  {Marchesini}, {Oesch}, {Pacifici}, {Patel}, {Price}, {Tal}, {Wake}, {van der
  Wel}, \& {Wuyts}}]{2016ApJS..225...27M}
{Momcheva}, I.~G., {Brammer}, G.~B., {van Dokkum}, P.~G., {et~al.} 2016, \apjs,
  225, 27, \dodoi{10.3847/0067-0049/225/2/27}

\bibitem[{{Murphy}(2013)}]{2013ApJ...777...58M}
{Murphy}, E.~J. 2013, \apj, 777, 58, \dodoi{10.1088/0004-637X/777/1/58}

\bibitem[{{Nelson} {et~al.}(2014){Nelson}, {van Dokkum}, {Franx}, {Brammer},
  {Momcheva}, {Schreiber}, {da Cunha}, {Tacconi}, {Bezanson}, {Kirkpatrick},
  {Leja}, {Rix}, {Skelton}, {van der Wel}, {Whitaker}, \&
  {Wuyts}}]{2014Natur.513..394N}
{Nelson}, E., {van Dokkum}, P., {Franx}, M., {et~al.} 2014, \nat, 513, 394,
  \dodoi{10.1038/nature13616}

\bibitem[{{Noeske} {et~al.}(2007){Noeske}, {Weiner}, {Faber}, {Papovich},
  {Koo}, {Somerville}, {Bundy}, {Conselice}, {Newman}, {Schiminovich}, {Le
  Floc'h}, {Coil}, {Rieke}, {Lotz}, {Primack}, {Barmby}, {Cooper}, {Davis},
  {Ellis}, {Fazio}, {Guhathakurta}, {Huang}, {Kassin}, {Martin}, {Phillips},
  {Rich}, {Small}, {Willmer}, \& {Wilson}}]{2007ApJ...660L..43N}
{Noeske}, K.~G., {Weiner}, B.~J., {Faber}, S.~M., {et~al.} 2007, \apjl, 660,
  L43, \dodoi{10.1086/517926}

\bibitem[{{O'Dea}(1998)}]{1998PASP..110..493O}
{O'Dea}, C.~P. 1998, \pasp, 110, 493, \dodoi{10.1086/316162}

\bibitem[{{Pettini} \& {Pagel}(2004)}]{2004MNRAS.348L..59P}
{Pettini}, M., \& {Pagel}, B.~E.~J. 2004, \mnras, 348, L59,
  \dodoi{10.1111/j.1365-2966.2004.07591.x}

\bibitem[{{Popping} {et~al.}(2017){Popping}, {Decarli}, {Man}, {Nelson},
  {B{\'e}thermin}, {De Breuck}, {Mainieri}, {van Dokkum}, {Gullberg}, {van
  Kampen}, {Spaans}, \& {Trager}}]{2017A&A...602A..11P}
{Popping}, G., {Decarli}, R., {Man}, A.~W.~S., {et~al.} 2017, \aap, 602, A11,
  \dodoi{10.1051/0004-6361/201730391}

\bibitem[{{Pound} \& {Wolfire}(2008)}]{2008ASPC..394..654P}
{Pound}, M.~W., \& {Wolfire}, M.~G. 2008, in Astronomical Society of the
  Pacific Conference Series, Vol. 394, Astronomical Data Analysis Software and
  Systems XVII, ed. R.~W. {Argyle}, P.~S. {Bunclark}, \& J.~R. {Lewis}, 654

\bibitem[{{Puglisi} {et~al.}(2019){Puglisi}, {Daddi}, {Liu}, {Bournaud},
  {Silverman}, {Circosta}, {Calabr{\`o}}, {Aravena}, {Cibinel}, {Dannerbauer},
  {Delvecchio}, {Elbaz}, {Gao}, {Gobat}, {Jin}, {Le Floc'h}, {Magdis},
  {Mancini}, {Riechers}, {Rodighiero}, {Sargent}, {Valentino}, \&
  {Zanisi}}]{2019arXiv190502958P}
{Puglisi}, A., {Daddi}, E., {Liu}, D., {et~al.} 2019, arXiv e-prints,
  arXiv:1905.02958.
\newblock \doarXiv{1905.02958}

\bibitem[{{Renaud} {et~al.}(2019){Renaud}, {Bournaud}, {Agertz}, {Kraljic},
  {Schinnerer}, {Bolatto}, {Daddi}, \& {Hughes}}]{2019A&A...625A..65R}
{Renaud}, F., {Bournaud}, F., {Agertz}, O., {et~al.} 2019, \aap, 625, A65,
  \dodoi{10.1051/0004-6361/201935222}

\bibitem[{{Ricciardelli} {et~al.}(2010){Ricciardelli}, {Trujillo}, {Buitrago},
  \& {Conselice}}]{2010MNRAS.406..230R}
{Ricciardelli}, E., {Trujillo}, I., {Buitrago}, F., \& {Conselice}, C.~J. 2010,
  \mnras, 406, 230, \dodoi{10.1111/j.1365-2966.2010.16693.x}

\bibitem[{{Rodighiero} {et~al.}(2015){Rodighiero}, {Brusa}, {Daddi},
  {Negrello}, {Mullaney}, {Delvecchio}, {Lutz}, {Renzini}, {Franceschini},
  {Baronchelli}, {Pozzi}, {Gruppioni}, {Strazzullo}, {Cimatti}, \&
  {Silverman}}]{2015ApJ...800L..10R}
{Rodighiero}, G., {Brusa}, M., {Daddi}, E., {et~al.} 2015, \apj, 800, L10,
  \dodoi{10.1088/2041-8205/800/1/L10}

\bibitem[{{Sargent} {et~al.}(2014){Sargent}, {Daddi}, {B{\'e}thermin},
  {Aussel}, {Magdis}, {Hwang}, {Juneau}, {Elbaz}, \& {da
  Cunha}}]{2014ApJ...793...19S}
{Sargent}, M.~T., {Daddi}, E., {B{\'e}thermin}, M., {et~al.} 2014, \apj, 793,
  19, \dodoi{10.1088/0004-637X/793/1/19}

\bibitem[{{Schawinski} {et~al.}(2010){Schawinski}, {Urry}, {Virani}, {Coppi},
  {Bamford}, {Treister}, {Lintott}, {Sarzi}, {Keel}, {Kaviraj}, {Cardamone},
  {Masters}, {Ross}, {Andreescu}, {Murray}, {Nichol}, {Raddick}, {Slosar},
  {Szalay}, {Thomas}, \& {Vandenberg}}]{2010ApJ...711..284S}
{Schawinski}, K., {Urry}, C.~M., {Virani}, S., {et~al.} 2010, \apj, 711, 284,
  \dodoi{10.1088/0004-637X/711/1/284}

\bibitem[{{Schinnerer} {et~al.}(2016){Schinnerer}, {Groves}, {Sargent},
  {Karim}, {Oesch}, {Magnelli}, {LeFevre}, {Tasca}, {Civano}, {Cassata}, \&
  {Smol{\v{c}}i{\'c}}}]{2016ApJ...833..112S}
{Schinnerer}, E., {Groves}, B., {Sargent}, M.~T., {et~al.} 2016, \apj, 833,
  112, \dodoi{10.3847/1538-4357/833/1/112}

\bibitem[{{Schmidt}(1959)}]{1959ApJ...129..243S}
{Schmidt}, M. 1959, \apj, 129, 243, \dodoi{10.1086/146614}

\bibitem[{{Schreiber} {et~al.}(2015){Schreiber}, {Pannella}, {Elbaz},
  {B{\'e}thermin}, {Inami}, {Dickinson}, {Magnelli}, {Wang}, {Aussel}, {Daddi},
  {Juneau}, {Shu}, {Sargent}, {Buat}, {Faber}, {Ferguson}, {Giavalisco},
  {Koekemoer}, {Magdis}, {Morrison}, {Papovich}, {Santini}, \&
  {Scott}}]{2015A&A...575A..74S}
{Schreiber}, C., {Pannella}, M., {Elbaz}, D., {et~al.} 2015, \aap, 575, A74,
  \dodoi{10.1051/0004-6361/201425017}

\bibitem[{{Scoville} {et~al.}(2007){Scoville}, {Aussel}, {Brusa}, {Capak},
  {Carollo}, {Elvis}, {Giavalisco}, {Guzzo}, {Hasinger}, {Impey}, {Kneib},
  {LeFevre}, {Lilly}, {Mobasher}, {Renzini}, {Rich}, {Sanders}, {Schinnerer},
  {Schminovich}, {Shopbell}, {Taniguchi}, \& {Tyson}}]{2007ApJS..172....1S}
{Scoville}, N., {Aussel}, H., {Brusa}, M., {et~al.} 2007, \apjs, 172, 1,
  \dodoi{10.1086/516585}

\bibitem[{{Scoville} {et~al.}(2014){Scoville}, {Aussel}, {Sheth}, {Scott},
  {Sanders}, {Ivison}, {Pope}, {Capak}, {Vanden Bout}, {Manohar}, {Kartaltepe},
  {Robertson}, \& {Lilly}}]{2014ApJ...783...84S}
{Scoville}, N., {Aussel}, H., {Sheth}, K., {et~al.} 2014, \apj, 783, 84,
  \dodoi{10.1088/0004-637X/783/2/84}

\bibitem[{{Scoville} {et~al.}(2016){Scoville}, {Sheth}, {Aussel}, {Vanden
  Bout}, {Capak}, {Bongiorno}, {Casey}, {Murchikova}, {Koda},
  {{\'A}lvarez-M{\'a}rquez}, {Lee}, {Laigle}, {McCracken}, {Ilbert}, {Pope},
  {Sanders}, {Chu}, {Toft}, {Ivison}, \& {Manohar}}]{2016ApJ...820...83S}
{Scoville}, N., {Sheth}, K., {Aussel}, H., {et~al.} 2016, \apj, 820, 83,
  \dodoi{10.3847/0004-637X/820/2/83}

\bibitem[{{Scoville} {et~al.}(2017){Scoville}, {Lee}, {Vanden Bout},
  {Diaz-Santos}, {Sanders}, {Darvish}, {Bongiorno}, {Casey}, {Murchikova},
  {Koda}, {Capak}, {Vlahakis}, {Ilbert}, {Sheth}, {Morokuma-Matsui}, {Ivison},
  {Aussel}, {Laigle}, {McCracken}, {Armus}, {Pope}, {Toft}, \&
  {Masters}}]{2017ApJ...837..150S}
{Scoville}, N., {Lee}, N., {Vanden Bout}, P., {et~al.} 2017, \apj, 837, 150,
  \dodoi{10.3847/1538-4357/aa61a0}

\bibitem[{{Skelton} {et~al.}(2014){Skelton}, {Whitaker}, {Momcheva}, {Brammer},
  {van Dokkum}, {Labb{\'e}}, {Franx}, {van der Wel}, {Bezanson}, {Da Cunha},
  {Fumagalli}, {F{\"o}rster Schreiber}, {Kriek}, {Leja}, {Lundgren}, {Magee},
  {Marchesini}, {Maseda}, {Nelson}, {Oesch}, {Pacifici}, {Patel}, {Price},
  {Rix}, {Tal}, {Wake}, \& {Wuyts}}]{2014ApJS..214...24S}
{Skelton}, R.~E., {Whitaker}, K.~E., {Momcheva}, I.~G., {et~al.} 2014, \apjs,
  214, 24, \dodoi{10.1088/0067-0049/214/2/24}

\bibitem[{{Speagle} {et~al.}(2014){Speagle}, {Steinhardt}, {Capak}, \&
  {Silverman}}]{2014ApJS..214...15S}
{Speagle}, J.~S., {Steinhardt}, C.~L., {Capak}, P.~L., \& {Silverman}, J.~D.
  2014, \apjs, 214, 15, \dodoi{10.1088/0067-0049/214/2/15}

\bibitem[{{Spilker} {et~al.}(2016){Spilker}, {Bezanson}, {Marrone}, {Weiner},
  {Whitaker}, \& {Williams}}]{2016ApJ...832...19S}
{Spilker}, J.~S., {Bezanson}, R., {Marrone}, D.~P., {et~al.} 2016, \apj, 832,
  19, \dodoi{10.3847/0004-637X/832/1/19}

\bibitem[{{Tacchella} {et~al.}(2016){Tacchella}, {Dekel}, {Carollo},
  {Ceverino}, {DeGraf}, {Lapiner}, {Mandelker}, \& {Primack
  Joel}}]{2016MNRAS.457.2790T}
{Tacchella}, S., {Dekel}, A., {Carollo}, C.~M., {et~al.} 2016, \mnras, 457,
  2790, \dodoi{10.1093/mnras/stw131}

\bibitem[{{Tacconi} {et~al.}(2008){Tacconi}, {Genzel}, {Smail}, {Neri},
  {Chapman}, {Ivison}, {Blain}, {Cox}, {Omont}, {Bertoldi}, {Greve},
  {F{\"o}rster Schreiber}, {Genel}, {Lutz}, {Swinbank}, {Shapley}, {Erb},
  {Cimatti}, {Daddi}, \& {Baker}}]{2008ApJ...680..246T}
{Tacconi}, L.~J., {Genzel}, R., {Smail}, I., {et~al.} 2008, \apj, 680, 246,
  \dodoi{10.1086/587168}

\bibitem[{{Tacconi} {et~al.}(2010){Tacconi}, {Genzel}, {Neri}, {Cox}, {Cooper},
  {Shapiro}, {Bolatto}, {Bouch{\'e}}, {Bournaud}, {Burkert}, {Combes},
  {Comerford}, {Davis}, {Schreiber}, {Garcia-Burillo}, {Gracia-Carpio}, {Lutz},
  {Naab}, {Omont}, {Shapley}, {Sternberg}, \& {Weiner}}]{2010Natur.463..781T}
{Tacconi}, L.~J., {Genzel}, R., {Neri}, R., {et~al.} 2010, \nat, 463, 781,
  \dodoi{10.1038/nature08773}

\bibitem[{{Tacconi} {et~al.}(2018){Tacconi}, {Genzel}, {Saintonge}, {Combes},
  {Garc{\'{\i}}a-Burillo}, {Neri}, {Bolatto}, {Contini}, {F{\"o}rster
  Schreiber}, {Lilly}, {Lutz}, {Wuyts}, {Accurso}, {Boissier}, {Boone},
  {Bouch{\'e}}, {Bournaud}, {Burkert}, {Carollo}, {Cooper}, {Cox}, {Feruglio},
  {Freundlich}, {Herrera-Camus}, {Juneau}, {Lippa}, {Naab}, {Renzini},
  {Salome}, {Sternberg}, {Tadaki}, {{\"U}bler}, {Walter}, {Weiner}, \&
  {Weiss}}]{2018ApJ...853..179T}
{Tacconi}, L.~J., {Genzel}, R., {Saintonge}, A., {et~al.} 2018, \apj, 853, 179,
  \dodoi{10.3847/1538-4357/aaa4b4}

\bibitem[{{Tadaki} {et~al.}(2017){Tadaki}, {Kodama}, {Nelson}, {Belli},
  {F{\"o}rster Schreiber}, {Genzel}, {Hayashi}, {Herrera-Camus}, {Koyama},
  {Lang}, {Lutz}, {Shimakawa}, {Tacconi}, {{\"U}bler}, {Wisnioski}, {Wuyts},
  {Hatsukade}, {Lippa}, {Nakanishi}, {Ikarashi}, {Kohno}, {Suzuki}, {Tamura},
  \& {Tanaka}}]{2017ApJ...841L..25T}
{Tadaki}, K.-i., {Kodama}, T., {Nelson}, E.~J., {et~al.} 2017, \apjl, 841, L25,
  \dodoi{10.3847/2041-8213/aa7338}

\bibitem[{{Tal} {et~al.}(2014){Tal}, {Dekel}, {Oesch}, {Muzzin}, {Brammer},
  {van Dokkum}, {Franx}, {Illingworth}, {Leja}, {Magee}, {Marchesini},
  {Momcheva}, {Nelson}, {Patel}, {Quadri}, {Rix}, {Skelton}, {Wake}, \&
  {Whitaker}}]{2014ApJ...789..164T}
{Tal}, T., {Dekel}, A., {Oesch}, P., {et~al.} 2014, \apj, 789, 164,
  \dodoi{10.1088/0004-637X/789/2/164}

\bibitem[{{Talia} {et~al.}(2018){Talia}, {Pozzi}, {Vallini}, {Cimatti},
  {Cassata}, {Fraternali}, {Brusa}, {Daddi}, {Delvecchio}, {Ibar}, {Liuzzo},
  {Vignali}, {Massardi}, {Zamorani}, {Gruppioni}, {Renzini}, {Mignoli},
  {Pozzetti}, \& {Rodighiero}}]{2018MNRAS.476.3956T}
{Talia}, M., {Pozzi}, F., {Vallini}, L., {et~al.} 2018, \mnras, 476, 3956,
  \dodoi{10.1093/mnras/sty481}

\bibitem[{{Thomson} {et~al.}(2014){Thomson}, {Ivison}, {Simpson}, {Swinbank},
  {Smail}, {Arumugam}, {Alexander}, {Beelen}, {Brandt}, {Chandra},
  {Dannerbauer}, {Greve}, {Hodge}, {Ibar}, {Karim}, {Murphy}, {Schinnerer},
  {Sirothia}, {Walter}, {Wardlow}, \& {van der Werf}}]{2014MNRAS.442..577T}
{Thomson}, A.~P., {Ivison}, R.~J., {Simpson}, J.~M., {et~al.} 2014, \mnras,
  442, 577, \dodoi{10.1093/mnras/stu839}

\bibitem[{{Tisani{\'c}} {et~al.}(2019){Tisani{\'c}}, {Smol{\v c}i{\'c}},
  {Delhaize}, {Novak}, {Intema}, {Delvecchio}, {Schinnerer}, {Zamorani},
  {Bondi}, \& {Vardoulaki}}]{2019A&A...621A.139T}
{Tisani{\'c}}, K., {Smol{\v c}i{\'c}}, V., {Delhaize}, J., {et~al.} 2019, \aap,
  621, A139, \dodoi{10.1051/0004-6361/201834002}

\bibitem[{{Toft} {et~al.}(2014){Toft}, {Smol{\v c}i{\'c}}, {Magnelli}, {Karim},
  {Zirm}, {Michalowski}, {Capak}, {Sheth}, {Schawinski}, {Krogager}, {Wuyts},
  {Sanders}, {Man}, {Lutz}, {Staguhn}, {Berta}, {Mccracken}, {Krpan}, \&
  {Riechers}}]{2014ApJ...782...68T}
{Toft}, S., {Smol{\v c}i{\'c}}, V., {Magnelli}, B., {et~al.} 2014, \apj, 782,
  68, \dodoi{10.1088/0004-637X/782/2/68}

\bibitem[{{Toft} {et~al.}(2017){Toft}, {Zabl}, {Richard}, {Gallazzi},
  {Zibetti}, {Prescott}, {Grillo}, {Man}, {Lee}, {G{\'o}mez-Guijarro},
  {Stockmann}, {Magdis}, \& {Steinhardt}}]{2017Natur.546..510T}
{Toft}, S., {Zabl}, J., {Richard}, J., {et~al.} 2017, \nat, 546, 510,
  \dodoi{10.1038/nature22388}

\bibitem[{{Valentino} {et~al.}(2018){Valentino}, {Magdis}, {Daddi}, {Liu},
  {Aravena}, {Bournaud}, {Cibinel}, {Cormier}, {Dickinson}, {Gao}, {Jin},
  {Juneau}, {Kartaltepe}, {Lee}, {Madden}, {Puglisi}, {Sanders}, \&
  {Silverman}}]{2018ApJ...869...27V}
{Valentino}, F., {Magdis}, G.~E., {Daddi}, E., {et~al.} 2018, \apj, 869, 27,
  \dodoi{10.3847/1538-4357/aaeb88}

\bibitem[{{van der Wel} {et~al.}(2012){van der Wel}, {Bell}, {H{\"a}ussler},
  {McGrath}, {Chang}, {Guo}, {McIntosh}, {Rix}, {Barden}, {Cheung}, {Faber},
  {Ferguson}, {Galametz}, {Grogin}, {Hartley}, {Kartaltepe}, {Kocevski},
  {Koekemoer}, {Lotz}, {Mozena}, {Peth}, \& {Peng}}]{2012ApJS..203...24V}
{van der Wel}, A., {Bell}, E.~F., {H{\"a}ussler}, B., {et~al.} 2012, \apjs,
  203, 24, \dodoi{10.1088/0067-0049/203/2/24}

\bibitem[{{van der Wel} {et~al.}(2014){van der Wel}, {Franx}, {van Dokkum},
  {Skelton}, {Momcheva}, {Whitaker}, {Brammer}, {Bell}, {Rix}, {Wuyts},
  {Ferguson}, {Holden}, {Barro}, {Koekemoer}, {Chang}, {McGrath},
  {H{\"a}ussler}, {Dekel}, {Behroozi}, {Fumagalli}, {Leja}, {Lundgren},
  {Maseda}, {Nelson}, {Wake}, {Patel}, {Labb{\'e}}, {Faber}, {Grogin}, \&
  {Kocevski}}]{2014ApJ...788...28V}
{van der Wel}, A., {Franx}, M., {van Dokkum}, P.~G., {et~al.} 2014, \apj, 788,
  28, \dodoi{10.1088/0004-637X/788/1/28}

\bibitem[{{van Dokkum} {et~al.}(2014){van Dokkum}, {Bezanson}, {van der Wel},
  {Nelson}, {Momcheva}, {Skelton}, {Whitaker}, {Brammer}, {Conroy},
  {F{\"o}rster Schreiber}, {Fumagalli}, {Kriek}, {Labb{\'e}}, {Leja},
  {Marchesini}, {Muzzin}, {Oesch}, \& {Wuyts}}]{2014ApJ...791...45V}
{van Dokkum}, P.~G., {Bezanson}, R., {van der Wel}, A., {et~al.} 2014, \apj,
  791, 45, \dodoi{10.1088/0004-637X/791/1/45}

\bibitem[{{van Dokkum} {et~al.}(2015){van Dokkum}, {Nelson}, {Franx}, {Oesch},
  {Momcheva}, {Brammer}, {F{\"o}rster Schreiber}, {Skelton}, {Whitaker}, {van
  der Wel}, {Bezanson}, {Fumagalli}, {Illingworth}, {Kriek}, {Leja}, \&
  {Wuyts}}]{2015ApJ...813...23V}
{van Dokkum}, P.~G., {Nelson}, E.~J., {Franx}, M., {et~al.} 2015, \apj, 813,
  23, \dodoi{10.1088/0004-637X/813/1/23}

\bibitem[{{Whitaker} {et~al.}(2012){Whitaker}, {van Dokkum}, {Brammer}, \&
  {Franx}}]{2012ApJ...754L..29W}
{Whitaker}, K.~E., {van Dokkum}, P.~G., {Brammer}, G., \& {Franx}, M. 2012,
  \apjl, 754, L29, \dodoi{10.1088/2041-8205/754/2/L29}

\bibitem[{{Whitaker} {et~al.}(2014){Whitaker}, {Franx}, {Leja}, {van Dokkum},
  {Henry}, {Skelton}, {Fumagalli}, {Momcheva}, {Brammer}, {Labb{\'e}},
  {Nelson}, \& {Rigby}}]{2014ApJ...795..104W}
{Whitaker}, K.~E., {Franx}, M., {Leja}, J., {et~al.} 2014, \apj, 795, 104,
  \dodoi{10.1088/0004-637X/795/2/104}

\bibitem[{{Whitaker} {et~al.}(2017){Whitaker}, {Bezanson}, {van Dokkum},
  {Franx}, {van der Wel}, {Brammer}, {F{\"o}rster-Schreiber}, {Giavalisco},
  {Labb{\'e}}, {Momcheva}, {Nelson}, \& {Skelton}}]{2017ApJ...838...19W}
{Whitaker}, K.~E., {Bezanson}, R., {van Dokkum}, P.~G., {et~al.} 2017, \apj,
  838, 19, \dodoi{10.3847/1538-4357/aa6258}

\bibitem[{{Williams} {et~al.}(2014){Williams}, {Giavalisco}, {Cassata},
  {Tundo}, {Wiklind}, {Guo}, {Lee}, {Barro}, {Wuyts}, {Bell}, {Conselice},
  {Dekel}, {Faber}, {Ferguson}, {Grogin}, {Hathi}, {Huang}, {Kocevski},
  {Koekemoer}, {Koo}, {Ravindranath}, \& {Salimbeni}}]{2014ApJ...780....1W}
{Williams}, C.~C., {Giavalisco}, M., {Cassata}, P., {et~al.} 2014, \apj, 780,
  1, \dodoi{10.1088/0004-637X/780/1/1}

\bibitem[{{Xue} {et~al.}(2016){Xue}, {Luo}, {Brandt}, {Alexander}, {Bauer},
  {Lehmer}, \& {Yang}}]{2016ApJS..224...15X}
{Xue}, Y.~Q., {Luo}, B., {Brandt}, W.~N., {et~al.} 2016, The Astrophysical
  Journal Supplement Series, 224, 15, \dodoi{10.3847/0067-0049/224/2/15}

\bibitem[{{Yun} {et~al.}(2001){Yun}, {Reddy}, \&
  {Condon}}]{2001ApJ...554..803Y}
{Yun}, M.~S., {Reddy}, N.~A., \& {Condon}, J.~J. 2001, \apj, 554, 803,
  \dodoi{10.1086/323145}

\bibitem[{{Zolotov} {et~al.}(2015){Zolotov}, {Dekel}, {Mandelker}, {Tweed},
  {Inoue}, {DeGraf}, {Ceverino}, {Primack}, {Barro}, \&
  {Faber}}]{2015MNRAS.450.2327Z}
{Zolotov}, A., {Dekel}, A., {Mandelker}, N., {et~al.} 2015, \mnras, 450, 2327,
  \dodoi{10.1093/mnras/stv740}

\end{thebibliography}

\end{document}